\documentclass[floatfix,aps,prb,reprint,showpacs,superscriptaddress]{revtex4-2}
\usepackage[left=1.5cm,right=1.5cm,top=2cm,bottom=2cm]{geometry}
\usepackage{xcolor}
\usepackage{graphicx,amsmath,wasysym}% Include figure files
\usepackage{dcolumn}% Align table columns on decimal point
\usepackage{bm}% bold math

\usepackage[colorlinks=true,linkcolor=blue,allcolors=blue]{hyperref}%
\usepackage{natbib}
\usepackage{geometry} 
\usepackage{ifthen}
\usepackage{slashed}
\usepackage{graphicx}
\usepackage{amssymb}
\begin{document}
\title{Jordan-Wigner fermionization of quantum spin systems on arbitrary 2D lattices: A mutual Chern-Simons approach}% Force line breaks with \\
\author{Jagannath Das}
\email{jagannath.das@theory.tifr.res.in}
\author{Aman Kumar}
\email{aman.kumar@tifr.res.in}
\author{Avijit Maity}
\email{avijit.maity@tifr.res.in}
\author{Vikram Tripathi}

\email{vtripathi@theory.tifr.res.in}
\affiliation{Department of Theoretical Physics, Tata Institute of
Fundamental Research, Homi Bhabha Road, Colaba, Mumbai 400005, India.}


\begin{abstract}
A variety of analytical approaches have been developed for the study of quantum spin systems in two dimensions, the notable ones being spin-waves, slave boson/fermion parton constructions, and for lattices with one-to-one local correspondence of faces and vertices, the 2D Jordan-Wigner (JW) fermionization. Field-theoretically, JW fermionization is implemented through Chern-Simons (CS) flux attachment. For a correct fermionization of lattice quantum spin-$1/2$ magnets, it is necessary that the fermions obey mutual bosonic (anyonic) statistics under exchange - this is not possible to implement on arbitrary 2D lattices if fermionic matter couples only to the lattice gauge fields. Enlarging the gauge degrees of freedom to include the dual lattice allows the construction of consistent mutual Chern-Simons field theories. Here we propose a mutual CS theory where the microscopic (spin) degrees of freedom are represented as lattice fermionic matter additionally coupled to specific combinations of dual lattice gauge fields that depend on the local geometry. We illustrate the use of this method for understanding the properties of a honeycomb Kitaev model subjected to a strong Zeeman field in the $z$-direction. Our CS gauge theory framework provides an understanding why the topological phase is degraded at lower (higher) critical fields for the ferro- (antiferro-) magnetic Kitaev interaction. Additionally, we observe an effectively one-dimensional character of the low-excitations at higher fields in the $z$-direction which we also confirm by spin-wave calculations.
\end{abstract}
\maketitle
\section{Introduction}
 The Jordan-Wigner approach is very attractive for studying quantum spin-$1/2$ systems in two dimensions. Unlike the Holstein-Primakoff \cite{holstein1940field} (or interacting spin-wave) approaches, it does not generate highly nonlinear many-body interactions. Compared to parton-based approaches, it does not require enlarging the Hilbert space which then must be projected to the physical space \cite{wen2002quantum,teng2020unquantized,wang2006spin,knolle2018dynamics}. Moreover, the JW fermions naturally interpolate between magnons and spinons, as is evident in the study of a simpler 1D system - the transverse field Ising model - and readily describe fractionalized quasiparticles in different phases \cite{kitaev2010topological,kitaev2006anyons}. JW fermionization, from the outset, gives a topological (CS) field theory \cite{fradkin1989prl,wang1991ground,azzouz1993interchain}, where CS flux attachment generates the  interaction of the JW fermions. Chern-Simons field theories provide a natural language \cite{tong2016lectures} for describing topological phases and their emergent excitations \cite{subir2009prb,wen2008prb,fradkin2008dhep,tong2016prx}. CS flux attachment is easy to implement on lattices where a local association of every lattice site (vertex) with a unique face \cite{fradkin2015prb} exists. In arbitrary 2D lattices $L$ where local face-vertex correspondence may not always be there, consistent CS field theories may still be constructed by including gauge fields on dual lattice ($L^{*}$) sites and links. The result is a mutual Chern-Simons theory \cite{sen2000pre,subir2009prb,geraedts2012monte}, where every lattice site is locally associated with the (unique) face dual to the site. A complicating factor is that the mutual CS theory describes the mutual anyonic statistics of particles respectively living on $L$ and $L^{*}.$ For the JW fermions to describe quantum spins, we require an implementation of bosonic statistics for the exchange of fermionic matter on $L$ although in general the CS theory describes anyons. Here we propose a mutual CS theory where the microscopic spin degrees of freedom on $L$ are represented as lattice fermions living on $L$ attached to a certain local combination of dual lattice gauge fields living on $L^{*}$ such that the desired anyonic statistics is realized.
 
 The Chern-Simons formulation for spin lattices with face-vertex correspondence nevertheless suffers from some limitations. Even in the absence of, say, an external magnetic field, these CS theories are not parity and time-reversal invariant, unlike the original microscopic models \cite{susskind1991}. Besides, quantizing the field theory requires a careful handling of the lattice analogue of the Levi-Civita symbol \cite{fradkin2015prb}. Likewise, there is ambiguity in the commutation relation of two Wilson loops, one of which ends on the path of the other, unless one introduces a dual curve \cite{fradkin2015prb}. Mutual CS theories do not suffer from these shortcomings. They can even be formulated for arbitrary 2D lattices. For the special case of lattices with face-vertex correspondence, it was shown in Ref. \cite{susskind1991} that anyons may be represented in the mutual CS theory as extended (dumb-bell) fermionic fields whose ends live respectively on $L$ and nearest dual lattice $L^{*}$ sites. The continuum limit of this lattice theory describes point-like anyons. An alternate proposal made recently \cite{zhang2021,banks2021} involves starting with a mutual CS theory but imposing an additional constraint on the lattice (link) gauge fields that they should be equal to the average of gauge fields on the nearby dual links. The idea is to avoid the problems in the formulation of Ref. \cite{fradkin2015prb}, although when applied to hexagonal or triangular lattices, the formalism gives unphysical fractional values for the linking numbers of Wilson loops. In contrast to field-theory approaches based on Chern-Simons flux attachment, Hamiltonian approaches using the 2D Jordan-Wigner transformation have also been used for lattices including those lacking face-vertex correspondence \cite{Kamenevprb2020,kamenev2017prb}; however long-range interactions are generated in this process, and it is also unclear if large gauge fluctuations necessary for charge quantization are accounted for. 
 
 As an illustration of our proposed technique, we study the honeycomb Kitaev model in a strong Zeeman field ($h$) in the $z$-direction. At low fields, the model is known to describe a deconfined phase with long-range topological order characterized by a four-fold degenerate ground state on the torus, and fractionalized excitations in the form of free Majorana fermions and gapped $Z_2$ visons. For the ferromagnetic sign of the Kitaev interaction, the topological order is quite fragile, vanishing at Zeeman fields a few per cent of the Kitaev interaction \cite{kumar2022absence}. For antiferromagnetic Kitaev interactions, topological order persists to larger Zeeman fields, around a fifth of the Kitaev interaction. Recently, there is great interest in understanding if fractionalization and other signatures of topological order such as a half-quantized thermal Hall conductivity can re-emerge at sufficiently high fields in Kitaev materials whose ground state otherwise has long-range magnetic order \cite{zhu2018robust,gordon2019theory}. Even if topological order may be strictly speaking not present at such fields, it is important to understand how much of the properties could be understood from the point of view of gauge field fluctuations coupling to fractionalized matter. At very high fields, the ground state is a fully polarized paramagnet, and it would normally make sense to approach this regime using the Holstein-Primakoff transformation (spin-wave theory). However as the field is decreased, it is known that interactions of the spin waves become rapidly very important, and spin waves do not provide a good description at lower fields where topological order is about to get restored. This encourages us to take the CS approach and check its advantages and limitations. 
 
 We obtain an effective mutual Maxwell-Chern-Simons field theory coupled to a superfluid order parameter field - i.e. a gauged superfluid. We show how the parameters in this theory can be systematically obtained from the underlying microscopic ones. Starting from the high-field side, which in our formalism corresponds to a confined phase, we progressively decrease the field, identifying the onset of local superfluidity and eventually the establishment of a global superfluid phase through the suppression of vison fluctuations. This represents the transition to the topologically ordered phase. Our perturbative approach in inverse of the field strength prevents us from accessing the low-field Kitaev dominated regime. Near to the topological transition, we also study the possibility of vison dispersion \cite{song2022translation} and make a comparison with understanding obtained from perturbative studies from the low-field side \cite{joy2021dynamics}. 
 
 The rest of the paper is organized as follows. In Section \ref{sec:2} we introduce the lattice version of mutual CS gauge theory for lattices lacking local face-vertex correspondence, focusing on the example of a honeycomb lattice. We propose in Sec. \ref{sec:3} a way of realizing the required anyonic exchange statistics of the JW fermions by attaching a certain combination of dual lattice gauge fields to fermionic matter on the lattice sites. Section \ref{sec:4} illustrates an application of this mutual CS formulation for the honeycomb Kitaev model in a finite Zeeman field in the $z$-direction. Here we describe phases of the Kitaev model in the lattice gauge theory language, starting from the high field limit. We develop an understanding of the evolution of parameters in the effective field theory in terms of the original microscopic parameters. We conclude with a summary of our findings and a discussion in Sec. \ref{sec:5}.
\section{Mutual CS Gauge Theory on Lattices Lacking Face-Vertex Correspondence}\label{sec:2}

Consistent formulation of CS theories on the lattice requires that every vertex is attached to the flux through a unique plaquette, which is evidently possible when there is a local face-vertex correspondence \cite{fradkin2015prb}. In such cases, the Euclidean time CS action has the form 
\begin{equation}
    S=-\frac{i \kappa}{2\pi}\int d\tau [A_{v}M_{v,f}\phi_{f}-\cfrac{
    1}{2} A_{e}K_{e,e'} \dot{A_{e'}}].
    \label{eq:fradkin_CS}
\end{equation}
Here the repeated indices are summed over. The indices $v,f,e$ run over all vertices, faces and edges respectively. $A_{v}$ and $A_{e}$ are respectively the temporal and spatial components of the gauge fields, with the former associated with the sites and the latter with the links. $\phi_{f}$ is the flux  through the face $f$ associated with the vertex $v$ via face-vertex correspondence. The detailed description of the  matrices $M_{v,f}$ and $K_{e,e'}$ are not important for the purposes of this paper and can be found in Ref. \cite{fradkin2015prb}. $M_{v,f}$ dictates the flux attachment and $K_{e,e'}$ is the lattice analog of the Levi-Civita symbol. The canonical commutation relation is
\begin{equation}
    [A_{e},A_{e'}]=-\frac{2\pi i}{\kappa} K^{-1}_{e,e'}.
    \label{eq:fradkin_comm}
\end{equation}
The $K_{e,e'}$ matrix in Eq. (\ref{eq:fradkin_comm}) is in general quite complicated and involves both forward and backward (spatial) differences \cite{semenoff1992prb}. It is also not very local in the sense that $e$ and $e'$ merely need to be associated with the same face
In case of lattices without face-vertex correspondence this $K_{e,e'}$ matrix is singular and the CS theory is no longer  consistent \cite{fradkin2015prb}. These difficulties are not due to some fundamental obstruction to defining lattice CS theories on arbitrary cellulations, since it should be possible to recover the continuum CS theory as a limiting case of any lattice. For quantum spin-$1/2$ lattice systems that we are ultimately interested in, we note that the Hamiltonian equivalent of CS theory - the 2D Jordan-Wigner transformations - do not have any requirement that the lattice must have local face-vertex correspondence. Such an approach has been taken, for example, for the XY model on the honeycomb lattice \cite{kamenev2017prb}.

\begin{figure}
\centering
\includegraphics[width=\columnwidth]{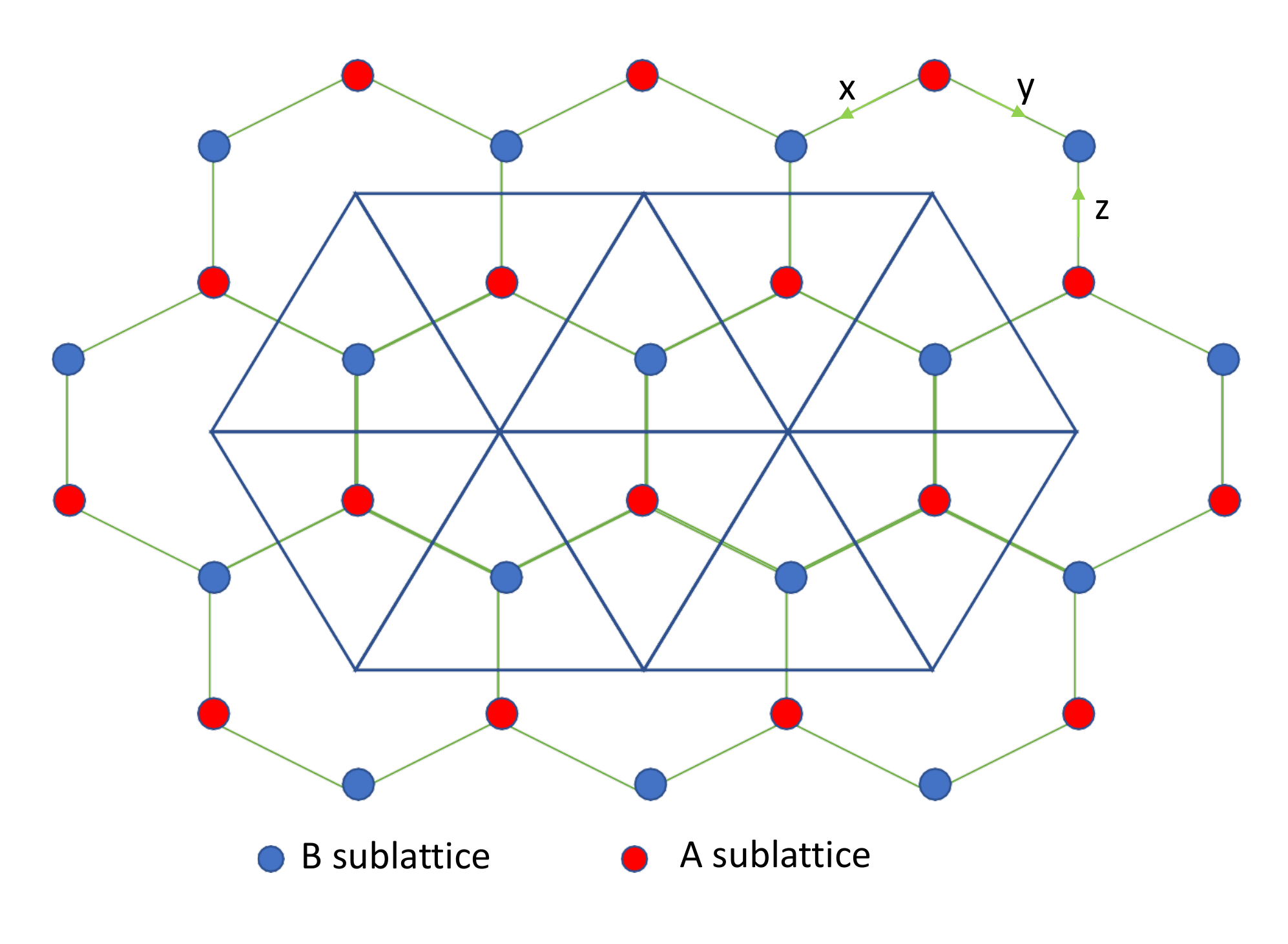}
\caption{Honeycomb lattice ($L$) and its dual triangular lattice ($L^{*}$). The faces of the triangular lattice are dual to the honeycomb lattice vertices, and likewise the hexagonal plaquettes are dual to the vertices of the triangular lattice. The dual of a link on $L$ is the link on $L^{*}$ crossing perpendicularly the relevant link on $L.$}
\label{fig:honeycomb}
\end{figure}

 Later in this paper, we will study as an example the honeycomb Kitaev model whose lattice evidently does not satisfy local face-vertex correspondence. The dual lattice is triangular, so the combined system has an equal number of vertices and faces, and moreover has local face-vertex  correspondence. It is easily seen that such local face vertex correspondence exists for arbitrary polygonal cellulations of 2D space. Figure \ref{fig:honeycomb} shows a honeycomb lattice (green) and its dual triangular (blue) lattice.  
 We now describe a mutual CS theory consisting of gauge fields on both honeycomb $L$ and triangular  $L^{*}$. Denote the temporal and spatial components of the gauge field on $L$  by $A_{v}$ and $A_{e}$ respectively, and on $L^{*}$ by $a^{*}_{v^{*}}$ and $a^{*}_{e^{*}}$ respectively. Analogously to Eq. (\ref{eq:fradkin_CS}), we associate the scalar potential at any vertex (whether on $L$ or $L^{*}$) with the flux through the dual plaquette corresponding to the vertex. The $U(1)$ gauge invariant Lagrangian \cite{fradkin2015prb} satisfying the above flux attachment rules is given by
\begin{align}
\mathcal{L}_{\text{CS}} & =\frac{\kappa}{4\pi}[\xi^{*}_{f^{*}e^{*}} a^{*}_{e^{*}}A_{v}\delta_{f^{*}v}+D^{*}_{v^{*}e^{*}}a^{*}_{v^{*}}A_{e} \delta_{ee^{*}}-\partial_{0}a^{*}_{e^{*}}A_{e}\delta_{ee^{*}}] \nonumber \\
& -\frac{\kappa}{4\pi}[\xi_{fe} A_{e} a^{*}_{v^{*}}\delta_{fv^{*}}+D_{ve}A_{v}a^{*}_{e^{*}} \delta_{ee^{*}}-\partial_{0}A_{e}a^{*}_{e^{*}}\delta_{ee^{*}}].
\label{eq:mixed_CS}
\end{align}
 The $\xi_{fe}$ and $D_{ve}$ are respectively the lattice analogs of curl and gradient \cite{fradkin2015prb} operations. The $\delta$-functions are defined as follows: $\delta_{e,e^{*}} = 1$ if $e$ and $e^{*}$ are links dual to each other, and zero otherwise, and similarly for $\delta_{fv^{*}}$ etc. The canonical equal time commutation relations are
 \begin{align}
  [A_{e},a^{*}_{e^{*}}] & = i \frac{2\pi}{\kappa} \delta_{e,e^{*}} \times {\rm sgn}(\vec{n_{e}} \times \vec{n_{e^{*}}}),\nonumber \\ 
   [A_{e}, A_{e'}] & = [a^{*}_{e^{*}}, a^{*}_{e'^{*}}] = 0.
   \label{eq:commutation_mCS}
 \end{align}
 Unlike the earlier formulation, there are no difficulties with the lattice version of the Levi-Civita term since $K_{e,e'^{*}}=\delta_{e,e'^{*}} \times{\rm sgn}(\vec{n_{e}} \times \vec{n_{e^{*}}}),$ and $A_{e}$, $a^{*}_{e^{*}}$ are perpendicular to each other like continuum case.  Furthermore, it can be shown that such a mutual Chern-Simons gauge theory has parity and time reversal symmetry \cite{freedman2004ap}. 
 Commutation relations of chains follow from the canonical commutation relations in Eq. (\ref{eq:commutation_mCS}) and the Baker-Hausdorff-Campbell formula: 
 \begin{align}
  \left[\int_{\mathcal{C}} A_e, \int_{\mathcal{C}^{*}}a^{*}_{e^{*}}\right] & = i\frac{2\pi}{\kappa}\nu[\mathcal{C},\mathcal{C}^{*}],
  \label{eq:chain}
 \end{align}
where $\nu[\mathcal{C},\mathcal{C}^{*}]$ is the difference of right handed and left handed intersections of chains $\mathcal{C}$ and $\mathcal{C}^{*}.$ Correspondingly, the relation between the respective Wilson lines will be
\begin{align}
 W_{\mathcal{C}}W_{\mathcal{C}^{*}} & = e^{-i\frac{2\pi}{\kappa}\nu[\mathcal{C},\mathcal{C}^{*}]}W_{\mathcal{C}^{*}}W_{\mathcal{C}}.
 \label{eq:wilson}
\end{align}
If source terms coupling the gauge fields to charge and current are now introduced, varying the action with respect to the temporal components of the gauge fields gives us the flux attachment constraints for physical states:
\begin{align}
\xi^{*}_{f^{*}e^{*}} a^{*}_{e^{*}} \equiv \Phi_{f^{*}} & = \frac{4\pi}{\kappa}Q_{v}, \nonumber \\
\xi_{fe} A_{e} \equiv \Phi_{f} & = \frac{4\pi}{\kappa}Q_{v^{*}}.
\label{eq:fv}
\end{align}
Here $\Phi_{f(f^{*})}$ is the flux associated with the face $f(f^{*})$ and $Q_{{v}(v^{*})}$ is the vertex charge in $L(L^{*})$.

If the system is subjected to toroidal boundary conditions, the spatial manifold has two holes that can be enclosed by non-contractible loops. The only nontrivial commutators are between pairs of (dual) non-contractible loops, drawn along two independent polar directions of the torus. In particular for $\kappa=2,$ the zero energy state can be labelled by the eigenvalues ($W=\pm1$), one for each independent non-contractible loop along the two polar directions, i.e., this state has a nontrivial four fold degeneracy associated with these nonlocal string operators.

\section{Quantum spin-1/2    particles on the honeycomb and triangular  lattice}\label{sec:3}
 The 2D Jordan-Wigner (JW) transformation expresses the spin raising (lowering) operators in terms of fermion creation (annihilation) operators attached to an infinite string, essentially a disorder operator, that implements bosonic commutation relations between spins at different spatial sites by ensuring odd values of the paths' linking number $\nu[\mathcal{C},\mathcal{C}^{*}]$ when they are exchanged using arbitrary paths. The 2D disorder operator, unlike its 1D counterpart, is not unique \cite{fradkin1989prl,wang1991ground,azzouz1993interchain,derzhko2001jordan}, and the only purpose is to implement the spin statistics. However, it is readily constructed for arbitrary polygonal cellulations of the 2D space. For a path integral (CS) formulation, one needs suitable disorder operators defined in terms of the gauge fields on the links, which are attached to the fermion (matter) fields, and it is also very desirable for computational simplicity that the Hamiltonian involves only local combinations of the link fields. For example, in the widely studied $XY$ spin models on lattices with face-vertex correspondence (e.g. square or Kagome), the gauge fields coupling to the hopping fermions are simply the Wilson lines associated with the corresponding links \cite{lopez1994chern,kumar2014chern}.  
 
\begin{figure}
\centering
\includegraphics[width=\columnwidth]{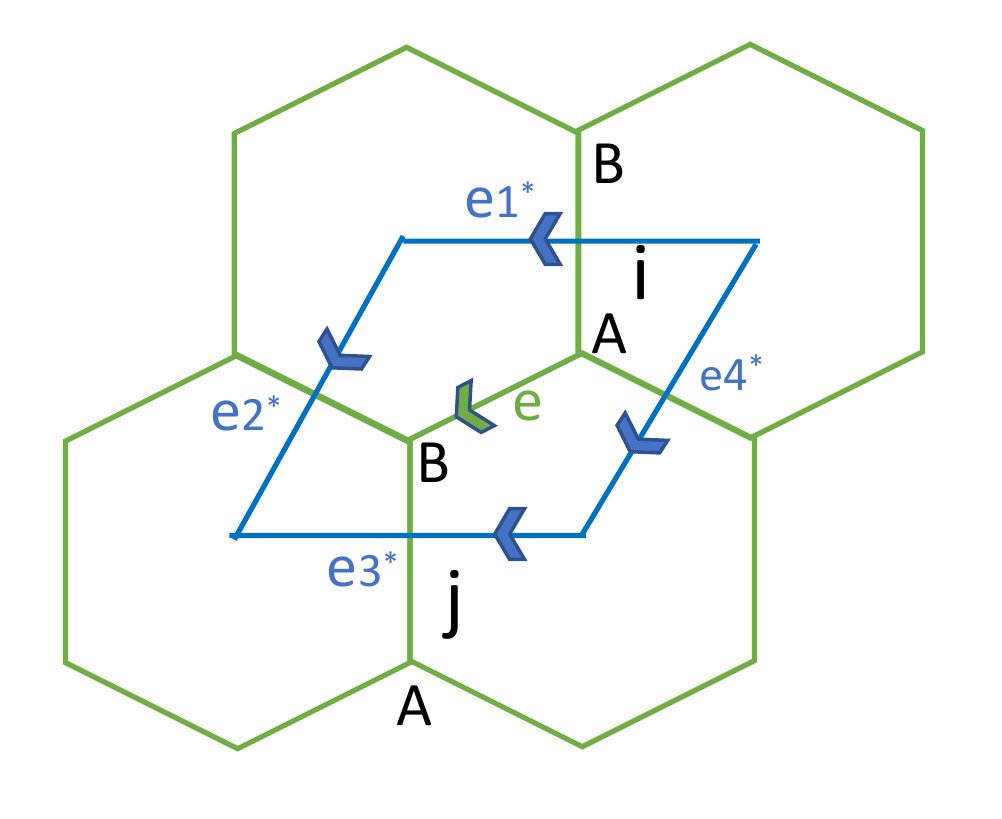}
\caption{\label{fig:JWCS}Anyonic (bosonic) exchange statistics of JW fermions on $L$ sharing link $e=i_{A}\rightarrow j_{B}$ is implemented in the mutual CS formulation (see text for details) by attaching the fermionic bilinear to a holonomy $e^{iB_e},$ where $B_e$ is proportional to the sum of the four dual link potentials on $L^{*}.$ Here $i,j$ are unit cells of the honeycomb lattice.}
\end{figure}

 Consider now spin-$1/2$ models on a lattice lacking face vertex correspondence - we take the honeycomb lattice first for concreteness. For simplicity, we are interested in models with only local couplings of fermions and gauge fields. One way to ensure this is by restricting ourselves to local Hamiltonians and  preserving fermion number parity. In Fig. \ref{fig:JWCS}, we show a schematic of a fermion bilinear sharing a link $(i_A, j_B)$ and attached to a local combination of lattice gauge fields, 
 \begin{align}
  \Psi_{i_A}^{\dagger} \Psi_{j_B} e^{iB_e} e^{iA_e},
  \label{eq:flux_att}
 \end{align}
where $B_e$ is a suitable local combination of the dual lattice gauge fields that we want to obtain. Such a term arises, for example, in two-body interaction of spins sharing a link. The choice of sign of the link gauge fields is chosen such that the holonomy $e^{iA_e}$ is associated for a hopping from site $j_B$ to site $i_A.$ We also need to give an orientation to our (directed) links - for our hexagonal lattice, the links are oriented from  the $A$ to $B$ sublattice. The dual triangular lattice is not bipartite, but here the orientation of the dual link is chosen such that the sign of $(\vec{n_{e}} \times \vec{n_{e^{*}}})$ is positive. Since the process in Eq. (\ref{eq:flux_att}) conserves fermion number, $A_e$ can be in general $U(1).$

It is important to note that since we have a mutual CS theory, the link fields $A_e$ are not responsible for anyonic (bosonic) statistics of exchange of spins on different lattice sites, and such statistics comes entirely from attaching our lattice fermion fields to the dual lattice gauge fields. The choice of $B_e$ is not unique. We propose (see Fig. \ref{fig:JWCS})
\begin{align}
 B_{e} & = N (a^{*}_{e1^{*}}+a^{*}_{e2^{*}}+a^{*}_{e3^{*}}+a^{*}_{e4^{*}}),
 \label{eq:B}
\end{align}
to be the sum of the four fields in the rhombus enclosing the link up to some normalisation constant $N.$ For our lattice, we argue that the normalization $N=\frac{\kappa}{4}$ ensures the desired statistics.

\begin{figure}
\centering
\includegraphics[width=\columnwidth]{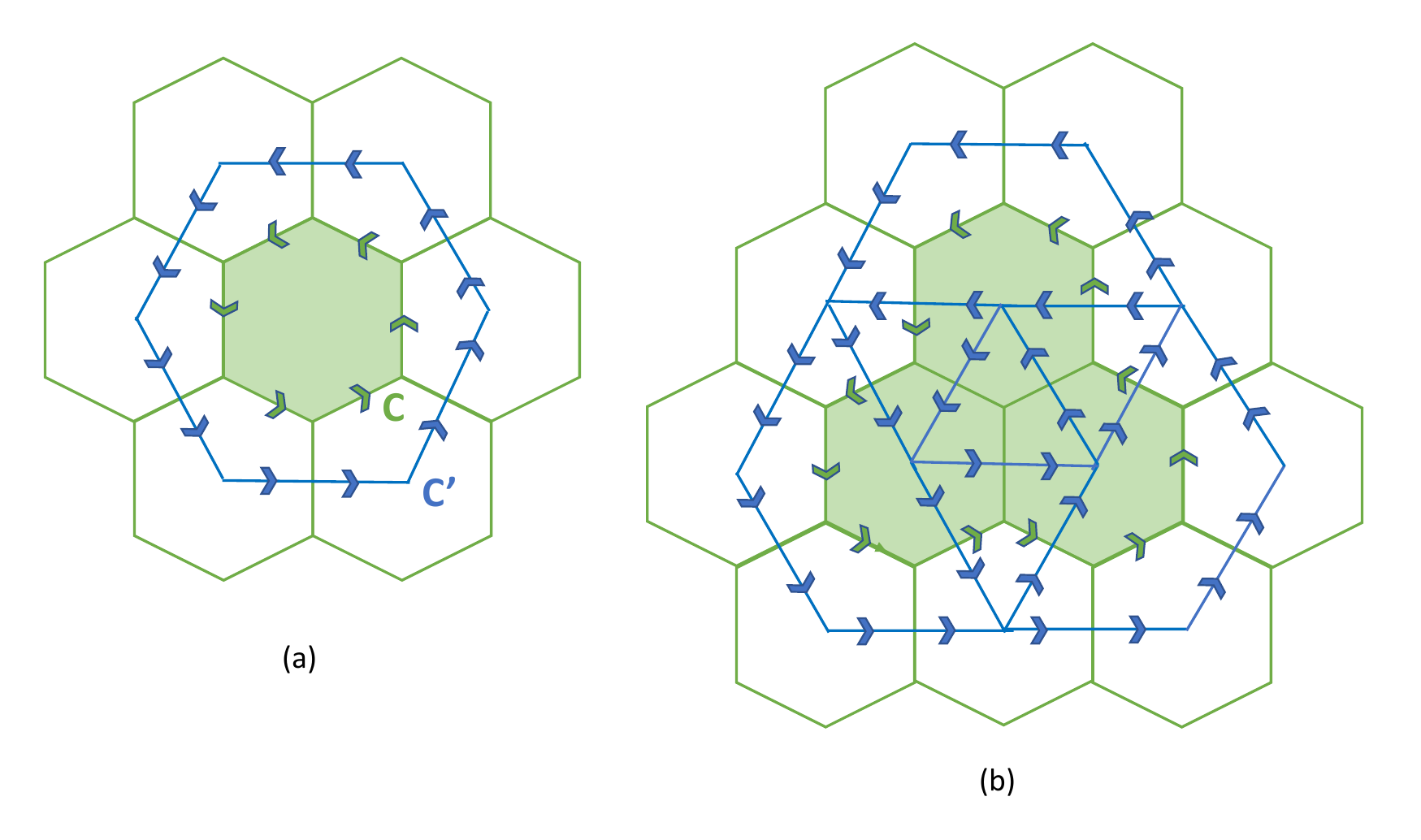}
\caption{\label{fig:N} Schematic of lattice and dual loops traversed upon taking a spin-$1/2$ particle around two different loops on the honeycomb lattice, denoted by the boundaries of the shaded regions. The arrows represent the number of times the corresponding edge is traversed. Note that each dual loop makes two windings, which in turn determines the normalization of $B_e.$ }
\end{figure}

Let us first take a spin-$1/2$ particle around the elementary hexagonal plaquette on the honeycomb lattice (see Fig. \ref{fig:N}a), which results in one $2\pi$ winding of the hexagonal void, and a double winding of the dual lattice links enclosing the vertices of the hexagon. On the dual lattice path, the accumulated phase is 
\begin{align}
 2 N \xi^{*}_{f^{*}e^{*}} a^{*}_{e^{*}}  & =\frac{8 N \pi}{\kappa}  \sum_{v} Q_{v}.
 \label{eq:hex_loop}
\end{align}
Since there are no lattice charges enclosed by the dual lattice path in this case, the phase is zero, and $N$ cannot be fixed here. Now let us take the spin-$1/2$ particle around the simplest (3-hexagon) path that encloses a vertex (see Fig. \ref{fig:N}b). Here the phase accumulated by the dual curve is  
\begin{align}
 3 \times 2 N \xi^{*}_{f^{*}e^{*}} a^{*}_{e^{*}}  & = \frac{24 N \pi Q_{v}}{\kappa},
 \label{eq:3hex_loop}
\end{align}
with $Q_v=1.$ Correct spin statistics requires an odd multiple of $2\pi$ accumulated by the dual loops, the simplest choice at first sight appears to be $N=\kappa/12.$ However with such a choice, it is known \cite{zhang2021} that intersecting Wilson loops have unphysical $1/3$ (fractional) linking number. This is also evident from the dual path in Fig. \ref{fig:N}a, where $\sum_v Q_v = 2,$ which is equivalent to $\sum_v Q_v=0,$ results in a fractional phase of $2\pi/3$ per particle, instead of $2\pi.$ The problem is that the dual curve in the left figure has one $4\pi$ winding, while the one on the right has three $4\pi$ windings. Thus the correct values for $N$ are $N=\frac{(2m+1)}{4}\kappa,$ where $m$ is an integer. Without loss of generality, we choose $m=0,$ i.e., $N=\kappa/4.$ The vortex charges $Q_v^{*}$ do not depend on the normalization. Note that the lattice Wilson loops square to unity; accordingly, the vortex charges are integer multiples of $\kappa/4.$ 

 \begin{figure}[t]
\centering
\includegraphics[width=\columnwidth]{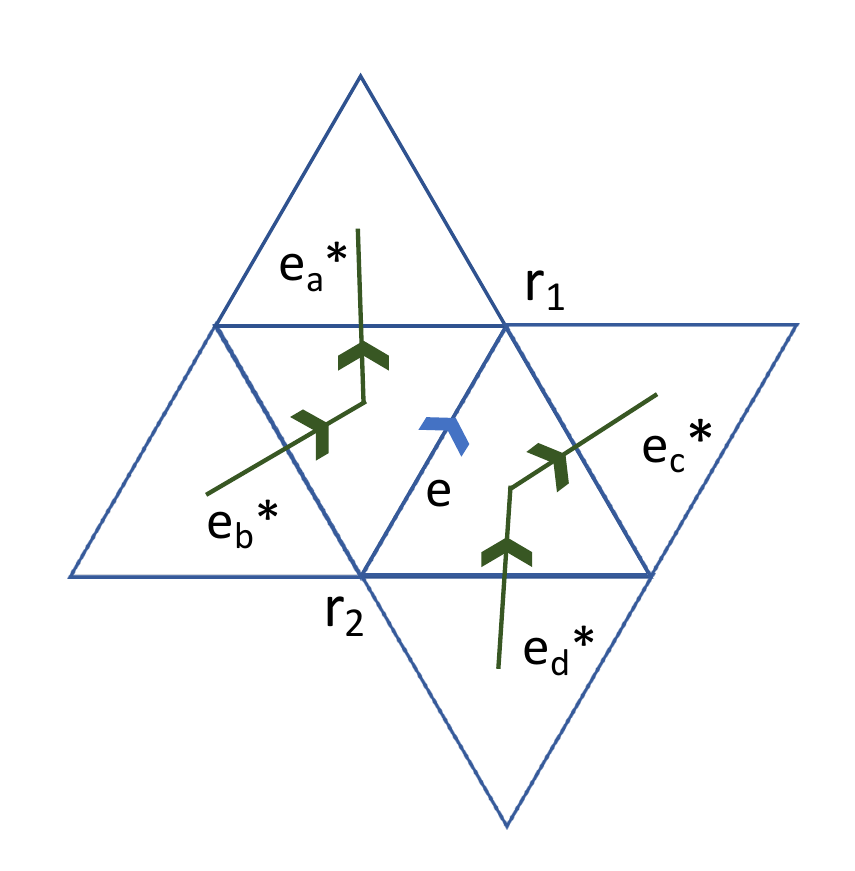}
\caption{\label{fig:tri_1}Arrangement of dual lattice gauge fields for a triangular lattice for implementing bosonic exchange statistics for the JW fermions sharing the link $e=r_{2}\rightarrow r_{1}.$ This lattice also lacks one-to-one face-vertex correspondence.}
\end{figure}
Now we use the same procedure for triangular lattice where $A_{e}$ and $a^*_{e^*}$ live on the direct triangular lattice edge and dual honeycomb lattice edge respectively. Here the choice of $B_{e}$ is (see Fig. \ref{fig:tri_1})

\begin{align}
 B_{e} & = N (a^{*}_{e^{*}_a}+a^{*}_{e^{*}_b}+a^{*}_{e^{*}_c}+a^{*}_{e^{*}_d})
 \label{eq:tri}.
\end{align}

\begin{figure}
\centering
\includegraphics[width=\columnwidth]{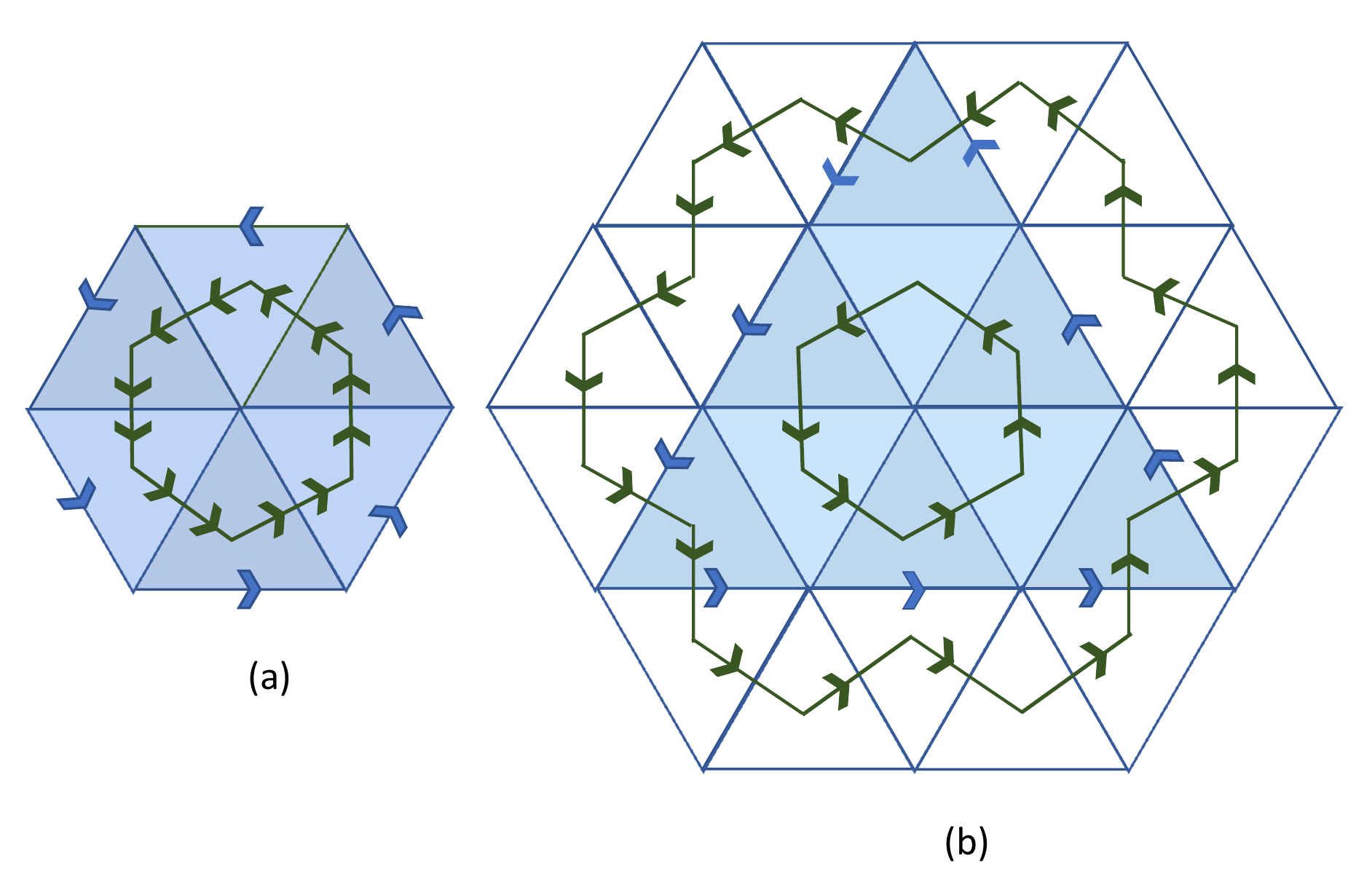}
\caption{\label{fig:N_tri} Schematic of lattice (blue) and dual (green) loops traversed upon taking a spin-$1/2$ particle around two different loops (boundary of the shaded regions) on the triangular lattice.}
\end{figure}
Taking a spin-$\frac{1}{2}$ particle around a lattice loop in Fig. \ref{fig:N_tri}a or in Fig. \ref{fig:N_tri}b, in both cases the phase accumulated by the dual curve  is 
\begin{align}
 2 N \xi^{*}_{f^{*}e^{*}} a^{*}_{e^{*}}  & =\frac{8 N \pi}{\kappa}  \sum_{v} Q_{v},
 \end{align}
with $Q_{v}=1$. This results the normalisation $N$ for triangular lattice to be $N=\frac{(2m+1)}{4}\kappa$ same as honeycomb lattice. Note that for the case of honeycomb lattice the above choice of $B_{e}$ is not unique . If we take all the eight contributing links (instead of taking only four links making the rhombus in Fig. (\ref{fig:JWCS})) that connect the two ends of the dual link($e^{*}$) of any given link($e$), the normalisation can be shown to be $\frac{\kappa}{2}$. For a lattice such as Kagome, we found that the choice is unique.
As we have discussed above, the role of the $B_e$ fields on the dual links is to implement bosonic statistics for exchange of spins. From Eq. (\ref{eq:fv}) it is clear that the dual fluxes can only take values $0$ or $4\pi$ since $Q_v$ can take values $0$ or $1.$ We choose these dual gauge fields to satisfy $U(1)$ symmetry, although other choices such as $Z_2$ can also be made. 

 We now discuss gauging another kind of fermion bilinear corresponding to a link Cooper pair $(\Psi_{i_A}^{\dagger} \Psi_{j_B}^{\dagger})$ that also appears in numerous spin models such as Ising or Kitaev, where $S_{z}$ is not a conserved quantity. This process creates a fermion pair sharing a link. Such terms couple to a pair of lattice Wilson lines that terminate at the end points of the links, i.e., $\Psi_{i_A}^{\dagger} \Psi_{j_B}^{\dagger}W_{C_{i_A}}W_{C_{j_{B}}} e^{-iB_e}.$ The Wilson line $W_{C_{i_{A}}}= \exp[-i\sum_{e'\in C_{i_{A}}} A_{e'}]$ on the lattice transports a fermion from the boundary at infinity to site $i_A$ along the string $C_{i_A}$, and similarly for $W_{C_{j_{B}}}.$ For periodic boundary conditions, the lines emanate from a fermion pair annihilation on a link, and end at the pair creation link. These gauge fields are $U(1)$ in general. If the lattice fermions are strongly gapped, such as when there is a large Zeeman field, the effective gauge theory obtained after integrating out the fermions will not involve long strings, in which case we will get a local effective $U(1)$ gauge theory. If however the $A_e$ are $Z_2,$ the product $ W_{C_{i_A}}W_{C_{j_{B}}}$ further reduces to the holonomy $e^{-iA_e}$ on the link,
\begin{align}
  W_{C_{i_A}}W_{C_{j_{B}}} & \equiv e^{-iA_e},\, A_e \in Z_2.
  \label{eq:flux_att2}
 \end{align}

 \section{Application to Kitaev model in a large Zeeman field}\label{sec:4}
Having described the construction of our CS theory (with fermionic matter) for quantum spin systems on lattices that lack face-vertex correspondence, we apply our ideas to the ferromagnetic Kitaev model on the honeycomb lattice subjected to a large magnetic field along the $z$-direction such that the Kitaev interactions can be regarded as a perturbation.  The Hamiltonian is given by \cite{kitaev2006anyons}
 \begin{align}
\mathcal{H} & = \mathcal{H}_{0} + \mathcal{H}_{1} \nonumber \\
    &  = h\sum_{p}\sigma^{z}_{p} - \sum_{\langle pq\rangle \in \gamma-\rm{links}} J_{\gamma}\sigma_{p}^{\gamma}\sigma_{q}^{\gamma},
\end{align}
where $\mathcal{H}_{0,1}$ respectively refer to the Zeeman  and Kitaev terms, the $\sigma$ are Pauli matrices, $p,q$ are the  vertices associated with the corresponding link $\gamma,$ with $\gamma = x,\,y,\,\mbox{or }z,$ and $h$ is the strength of applied field. The ground state of the unperturbed (purely Zeeman) model is trivially a fully polarized paramagnet, and we set $\kappa=1.$ As the Zeeman field is progressively decreased, the system ultimately transitions into a deconfined state with fractionalized excitations and long-range topological order. The fact that relatively small field values ($h/J < 1$) suffice to degrade the topological order motivates us to approach the problem from the high field side. Our approach is also an alternative to the spin-wave approximation often employed for studies of the Kitaev model at high fields \cite{holstein1940field,dyson1956general,oguchi1960theory}. 

 We fermionize our model using the dual CS formalism described in Sec.\ref{sec:3}.  Specifically, we will obtain an effective field theory in the limit of large Zeeman field, $h \gg J_{\gamma}$ treating the Kitaev interactions as a perturbation. In this high field limit, the effective field theory is $U(1)$ gauge-invariant. In the opposite limit of large Kitaev interactions, the fermion number is not conserved but the fermion number parity is - implying that $U(1)$ gauge symmetry will break down to $Z_2.$

 After fermionization, the Kitaev Hamiltonian takes the form
\begin{equation}
    \mathcal{H}=\mathcal{H}_{x}+\mathcal{H}_{y}+\mathcal{H}_{z}+h\sum_{p}[2\Psi^{\dagger}_{p}\Psi_{p}-1],
    \label{eq:h_fermi}
\end{equation}
where 
\begin{align}
\mathcal{H}_{x} & =-J_{x}\!\!\!\!\!\!\sum_{x-\text{links(e)}}\!\!\!\!\![\Psi^{\dagger}_{p} e^{-i(A_{e}+B_{e})}\Psi^{\dagger}_{q}+\Psi^{\dagger}_{p}e^{i(A_{e}+B_{e})}\Psi_{q}+ \rm{h.c.}], \nonumber \\
\mathcal{H}_{y} & =-J_{y}\!\!\!\!\!\!\sum_{y-\text{links(e)}}\!\!\!\!\![-\Psi^{\dagger}_{p}e^{-i(A_{e}+B_{e})}\Psi^{\dagger}_{q}+\Psi^{\dagger}_{p}e^{i(A_{e}+B_{e})}\Psi_{q}+ \rm{h.c.}],\nonumber \\
\mathcal{H}_{z} & =-J_{z}\!\!\!\!\!\!\sum_{z-\text{links(e)}}\!\!\!\!\![2\Psi^{\dagger}_{p}\Psi_{p}-1][2\Psi^{\dagger}_{q}\Psi_{q}-1].
\end{align}
Since the fermionic matter is gauged only under the lattice gauge fields, and there is no vortex matter at the dual sites, $\mathcal{H}_x$ and $\mathcal{H}_y$ are not invariant under $U(1)$ gauge transformations of the dual lattice gauge fields $B_{e}.$  However, we may regard the above coupling of $B_e$ to the fermionic matter as an interaction in which the dynamics of $B_e$ is governed by the CS term - in this way, $B_e$ is an external dynamical field coupling to the fermions for the purpose of satisfying correct spin statistics and gauge invariance is not necessary.

Unlike the magnetization (or fermion number), which is a conserved quantity at high fields, the vortex charge is not. However in the Kitaev limit $h/J\rightarrow 0,$ the vortex charge is conserved. The role of a small Zeeman perturbation is to create pairs of vortices along the $z$-bonds in each order of the perturbation - thus at high fields where vortex number is ill-defined, the vortex number parity is still conserved. This makes us choose $A_e$ in the rest of the paper to be $Z_2$ and not $U(1),$ although the analysis can be performed equally well with $A_e \in U(1).$

For our bipartite lattice, the vertices carry two labels, namely the unit cell $(i,j..),$ and the sub-lattice $(A,B).$ Motivated by the fact that in the Kitaev limit, the model is equivalent to a topological superconductor, we choose to decouple the four fermion interaction in the Cooper channel,
\begin{align}
\mathcal{H}_{z} & =-J_{z}\!\!\!\!\sum_{z-\text{links}}\!\Bigg\{1-2\Psi^{\dagger}_{i_A}\Psi_{i_A}-2\Psi^{\dagger}_{i_B}\Psi_{i_B}-4|\Delta_{i}|^2 \nonumber \\
 & +4\Delta^{*}_{i} \Psi_{i_A}\Psi_{i_B}-4\Delta_{i} \Psi^{\dagger}_{i_A}\Psi^{\dagger}_{i_B}\Bigg\},
\end{align}
where $\Delta_{i}=\langle\Psi_{i_A}\Psi_{i_B}\rangle$ is the  order parameter. An alternate choice of decoupling in the density channel was not pursued guided by the fact that the order parameter will be large in the presence of large Zeeman fields which is undesirable in a perturbative expansion for the free energy.  

We consider the Euclidean time action for this model,
\begin{equation}
  S[\Psi_p,\Delta] =\sum_{p}\int^{\beta}_{0} d\tau [\Psi_{p}^{\dagger}(\partial_{\tau}+iA_{p})\Psi_{p}-i\mathcal{L}_{\text{CS}}+\mathcal{H}],  
\end{equation}  
where $\mathcal{H}$ is the Hamiltonian after mean field decoupling of the $z$-link interactions.
Now at a high magnetic field, charges  $n_{p}$  i.e. $\Psi_p^\dagger \Psi_p$ have only small fluctuations, and consequently, $A_{p}$ can have large fluctuations. It is convenient to perform a gauge transformation to eliminate the strongly fluctuating potential fields that appear in the fermionic determinant through a gauge transformation of the fermionic fields,
 \begin{equation}
\Psi_{i_A}\rightarrow  \Psi_{i_A}e^{i\chi_{i_A}},
 \end{equation}
 and choosing $A_{i_A}=-\partial_{\tau} \chi_{i_A}.$ 
 The fermionic partition function is given by
 \begin{align}
 Z & =\!\!\!\int D{\text{(fields)}} e^{-S};\,\,
S=S_{0}+S_{c}+S'+S^{*}.\text{ Here}
\nonumber\\
& S_{0}=\sum_{p}\int_{\tau}\Psi^{\dagger}_{p}(\partial_{\tau}+\xi_{p})\Psi_{p},\hspace{0.2in} \xi_{p}=2h+2J_{z},\nonumber\\
& S_{c}=-\int_{\tau} [iL_{\text{CS}}-4J_{z}\sum_{i}|\Delta_{i}|^2].
\label{partition}
\end{align}
The remaining terms are Nambu off-diagonal, number nonconserving,
\begin{align}
& S^{*}=\int_{\tau} T_{0} =\sum_{i}\int_{\tau} T^{i}_{0}\hspace{0.15in}\text{and}\hspace{0.15in}S'=\int_{\tau} T,
\label{partition1}
 \end{align}
 where $T_{0}$ and $T$ are defined below. 
 \begin{align}
  T & =-J_{x}\sum_{x-\text{links}}[\Psi^{\dagger}_{i_A} e^{-i\varphi^{x}_{1}}\Psi^{\dagger}_{j_B}+\Psi^{\dagger}_{i_A} e^{i\varphi^{x}_{2}}\Psi_{j_B}+h.c.]\nonumber\\
  &-J_{y}\sum_{y-\text{links}}[-\Psi^{\dagger}_{i_A} e^{-i\varphi^{y}_{1}}\Psi^{\dagger}_{j_B}+\Psi^{\dagger}_{i_A} e^{i\varphi^{y}_{2}}\Psi_{j_B}+h.c.]
  \label{eq:T}
  \end{align}
 where $i_A\rightarrow j_B$ is a $x$ bond  or a $y$ bond of the honeycomb Kitaev model and 
 \begin{align}
 \varphi_{1} & =A_{e}+B_{e}+\chi_{i_A}+\chi_{j_B},\,\text{and} \nonumber \\ 
 \varphi_{2} & =A_{e}+B_{e}-\chi_{i_A}+\chi_{j_B}.
 \label{eq:phi-def}
 \end{align}
 The superscripts ($x$, $y$) on $\varphi_{1,2}$ in Eq. (\ref{eq:T}) refer to the type of link ($x$ or $y.$).
 The fermionic part of the action takes the form
 \begin{equation}
     S_{F}=S_{0}+S^{*}+S'=\Psi^{\dagger}G^{-1}\Psi,
 \end{equation}       
     where the inverse Green function is $G^{-1}=G^{-1}_{0}+T_{0}+T.$
 Here  $\Psi$ is  $4N$ component spinor in the Nambu notation ( $N$ is the number of unit cells in the honeycomb lattice) with $ \Psi^{\dagger}_{i}=\begin{bmatrix}
 \Psi_{i_A} &
 \Psi^{\dagger}_{i_A} &
 \Psi_{i_B} &
 \Psi^{\dagger}_{i_B} 
 \end{bmatrix}^{\dagger}.$ ${G^{i}_{0}}^{-1}$ and $T^i_0$ are shown below;  
 \begin{equation}
 {G^{i}_{0}}^{-1}\!\!\!\!=\frac{1}{2}\!\!
 \begin{bmatrix}
 \partial_{\tau}+\xi_{i_A} & 0 & 0 & 0 \\ \!
 0 & \partial_{\tau}-\xi_{i_A}  & 0 & 0 \\ \!
 0 & 0 & \partial_{\tau}+\xi_{i_B}  & 0 \\ \!
 0 & 0 & 0 & \partial_{\tau}-\xi_{i_B}  \\ \!
 \end{bmatrix},
 \label{eq:matrixG}
 \end{equation}
 \begin{equation*}
 T^{i}_{0}=2J_{z}
 \begin{bmatrix}
 0 & 0 & 0 & \Delta_{i}e^{-i\varphi^{i}_{1}} \\
 0 & 0  & -\Delta^{*}_{i}e^{i\varphi^{i}_{1}} & 0 \\
 0 & -\Delta_{i}e^{-i\varphi^{i}_{1}} & 0  & 0 \\
 \Delta^{*}_{i}e^{i\varphi^{i}_{1}} & 0 & 0 & 0  \\
 \end{bmatrix} 
 \end{equation*}
 where $\varphi^{i}_{1}=\chi_{i_A}+\chi_{i_B}$. The superscript $i$ in $\varphi^{i}_{1}$ runs over unit cell i.e. $z$-bond. We now formally integrate out the fermions (which are gapped in the presence of the strong Zeeman field),
 \begin{equation*}
     S=S_{c}-\text{tr}\ln(G^{-1}),\hspace{0.1in} \ln G^{-1}=\ln G^{-1}_{0} +\ln[1+G_{0}(T_{0}+T)],
 \end{equation*}
 and expand the logarithm in the small parameters $J/h,$
 \iffalse
 \begin{align*}
     \text{log}G^{-1} & =\text{log}G^{-1}_{0}+[G_{0}T_{0}+G_{0}T-\frac{1}{2}G_{0}T_{0}G_{0}T_{0} \\
    & -\frac{1}{2}G_{0}T_{0}G_{0}T -\frac{1}{2}G_{0}TG_{0}T_{0}-\frac{1}{2}G_{0}TG_{0}T+....].
 \end{align*}
 \fi
 We also drop $\ln G^{-1}_{0}$ as it does not involve any dynamical fields. In the expansion, the leading terms $\text{tr}(G_{0} T)$ and $\text{tr}(G_{0}T_{0})$ vanish because $G_{0}$ is site-diagonal and $T$, $T_{0}$ are site off-diagonal.  \iffalse Detailed expressions for the leading nonvanishing terms $\text{tr}(G_{0}T_{0}G_{0}T_{0})$ and $\text{tr}(G_{0}TG_{0}T)$ and other  higher orders are obtained in the  Appendix. \fi There are two types of terms that appear in the resulting effective field theory: link and loop terms, the leading contributions respectively appearing at the second and sixth order. Details of the derivation of the effective action are presented in the Appendix. Using these results we present our effective action,
 \begin{align}
    S & =S_{c}+\int_{\tau}\! \left[\sum_{x-links} \!\!\mathcal{L}^x_2+\!\!\!\!\sum_{y-links}\mathcal{L}^y_2+\!\!\!\!\sum_{z-links}\mathcal{L}^z_2+\sum_{\hexagon}\!\!\mathcal{L}_6\right].
    \label{eq:seff}
\end{align}
The leading contributions to the link terms at low temperatures ($\beta h \gg 1$) are:
\begin{align}
 \mathcal{L}^x_2 & =\frac{J^{2}_{x}}{64h^3} \left(\frac{\partial \varphi^{x}_{1}}{\partial \tau}-2ih\right)^2, \nonumber \\
  \mathcal{L}^y_2 & =\frac{J^{2}_{y}}{64h^3} \left(\frac{\partial \varphi^{y}_{1}}{\partial \tau}-2ih\right)^2, \nonumber \\
\mathcal{L}^z_{2} &
=\!\!\frac{-J^2_{z}}{2h^3} \!\!\left[ 16h^2|\Delta_i|^2\!\!-4h\Delta^{*}_i(\partial_{\tau}\!-i\dot{\varphi^i_1})\Delta_i\! 
-\!|(\partial_{\tau}-i\dot{\varphi^i_1})\Delta_i|^{2}\! \right].
\label{eq:L-links}
\end{align}
Contributions corresponding to $\varphi^{x(y)}_{2}$  are associated with particle-hole hopping and are proportional to $n_{F}(-\xi_{iA})n_{F}(\xi_{jB}),$ (see Appendix) which is negligible at high fields in the ground state sector (all sites completely empty or completely full), whereas $\varphi^{x(y)}_{1}$ terms are associated with particle-particle creation or annihilation on a link, and are proportional to $n_{F}(-\xi_{iA})n_{F}(-\xi_{jB})$. Here $n_{F}(\xi_{iA}/{\xi_{jB}})$ is the Fermi-Dirac distribution. Physically, the leading (second order in tunneling) particle-hole contribution to the effective action is forbidden by the Pauli exclusion. 
Particle-hole processes are relevant in higher energy sectors.

Consider now the definition of the phases $\varphi^{x(y)}_{1}$ in Eq. \ref{eq:phi-def}. Since the phases $\chi$ appearing in the definition of the $\varphi^{x(y)}_1$ are $Z_2,$ (i.e. taking values only $0$ or $\pi$) we can absorb them in a redefinition of the $A_e,$ which is equivalent to choosing a gauge where $A_v = 0.$ Thus the first two terms in Eq. \ref{eq:L-links} are reminiscent of the electric field terms in a Maxwell theory. The $i2h$ terms appearing with the electric field correspond to the Zeeman cost of flipping a spin. 

The loop term (that we refer to as a Josephson term because of its tendency to suppress the phase fluctuations) appears first only at the sixth order, 
\begin{align}
  \mathcal{L}_6= -\frac{8}{3} \int^{\beta}_{0} d\tau \frac{J^{2}_{x}J^{2}_{y}J^{2}_{z}}{(2h)^{5}}\Bigg\{\Delta_{j}\Delta^{*}_{l}e^{i \left[ \int_{\text {C}} \vec{A}.\vec{dl}+\int_{\text {C'}}{\frac{\vec{a^*}}{2}}.\vec{dl}\right]}+h.c.\Bigg\},
  \label{eq:wilson-6th}
\end{align}
where $C$ and $C'$ shown in Fig. \ref{fig:N} are respectively the hexagonal loop (with a single winding) on the lattice links, and the dual loop (with a double winding) that encloses this hexagon - both generated while taking a spin around the elementary hexagon. Since there is no fermion on the hexagon ($Q_v = 0$), the dual flux is zero, and without loss of generality, hereinafter we take the  order parameter fields to be real and its phase (i.e. sign) fluctuations are shifted to the gauge fields. As we have discussed earlier, the flux in $C$ can take values $0$ or $\pi.$ Fixing the signs of the order parameter fields to be the same, the energy associated with the plaquette terms is evidently minimized for $\int_{\text{C}} \vec{A}.\vec{dl} = \Phi_f =0.$ Equation (\ref{eq:seff}) describes a $Z_2$ gauged superfluid in which the dynamics of the gauge fields is governed by a mutual Maxwell-Chern-Simons theory. The Wilson loop term, Eq. (\ref{eq:wilson-6th}) determines the cost of a $\pi$-flux change in a plaquette (vison gap) - the cost clearly vanishes in the absence of superfluid order (i.e. when $\Delta=0$).

To simplify our further discussion, we limit ourselves to the isotropic Kitaev case i.e., $J_x = J_y = J_z = J.$ Upon reducing the field, the sign of the coefficient of the quadratic term ultimately turns negative, i.e.,
\begin{align}
 4J - 8 \frac{J^{2}}{h} & < 0,\,\mbox{or } h < 2J,
\end{align}
resulting in nonzero expectation values for local order parameter fields $\Delta_i.$ For the ferromagnetic Kitaev couplings ($J > 0$ in our model), a sufficiently small magnetic field is required for the $\Delta_i$ to develop a nonzero expectation. \iffalse Combining our high field result with the existing understanding that the Kitaev model in the low-field regime is essentially a gauged chiral p-wave superfluid, we posit that in the ferromagnetic case, there must be a spontaneous symmetry breaking transition of the superfluid order. \fi The situation is very different for the antiferromagnetic counterpart ($J < 0,$), where clearly $\Delta_i \neq 0 $ even at large Zeeman fields. This does not necessarily mean a superfluid phase for which, apart from a nonvanishing expectation for the local order parameter $\Delta$, we also need establishment of global phase coherence. Returning to our effective model, we note that the coupling constants for the Maxwell terms are $E_{C} = 16 h^3/J^2$ for the electric part and $E_{J} = J^6 |\Delta|^{2}/12 h^5$ for the magnetic part: 
\begin{align}
 \mathcal{L} & = \frac{1}{4 E_C}\!\!\!\sum_{e \in x,y \text{ links}} (\partial_{\tau}\varphi_{1}^{e} - 2ih)^{2} - E_J \sum_{\hexagon} \cos(\oint \vec{A}\cdot\vec{dl}).
\end{align}

The model has a dimensionless coupling constant, $g = E_{J}/E_{C} \sim (J/h)^8 \times (|\Delta|^{2}/ 192).$ When $g \gtrsim 1,$ the phase fluctuations of the $\Delta_i$ are suppressed resulting in the superfluid state. The superfluid phase here is associated with a broken $Z_2$ symmetry and not $U(1).$ This critical field from this criterion, \begin{align}
    g \gtrsim 1 \,\mbox{or } h\approx 0.518 J |\Delta(h)|^{1/4}
    \label{eq:transitionpoint}
\end{align}
is smaller compared to the field at which the local Cooper pairs are first formed. 

Consider now a large Wilson loop $W_{L}$ of perimeter $L$ that encloses a number $M \gg 1$ of elementary hexagonal plaquettes. 
At large magnetic field ($g\ll 1$), we perturbatively expand the exponential with the magnetic term and perform the average over the gauge field configurations. By Elitzur's theorem, only gauge invariant terms survive the averaging, and we get 
\begin{align}
 W_{L}  \sim \left(\frac{E_{J}}{E_{C}}\right)^{M},
 \label{eq:WL1}
\end{align}
where the angular brackets denote averaging over the order parameter field. Since $|\Delta|^2 = 0,$ we have $W_{L}\equiv 0,$ essentially a vortex superfluid which strongly confines the charges. Physically,  the large magnetic field suppresses spin flips or fermion number fluctuations. Conversely, large Wilson loops $W_{L^{*}}$ on the dual lattice are $O(1).$ Next, we reduce the magnetic field until $|\Delta|^2 \neq 0$ develops locally. If the local order parameter is small, we can still expand the exponential with the magnetic term. Clearly, the Wilson loop now follows a ``volume'' law,
\begin{align}
 W_{L} \sim \exp\left(-M \ln\left[\frac{E_{J}}{E_{C}}\right]\right),
 \label{eq:confine}
\end{align}
still indicating a confined phase. As our perturbative approach is valid only for $(h/J)^{8} > |\Delta|^{2}/192,$ we are unable to provide a complete description of the low field phase in the Kitaev limit.   

\begin{figure}[t]
\centering
\includegraphics[width=\columnwidth]{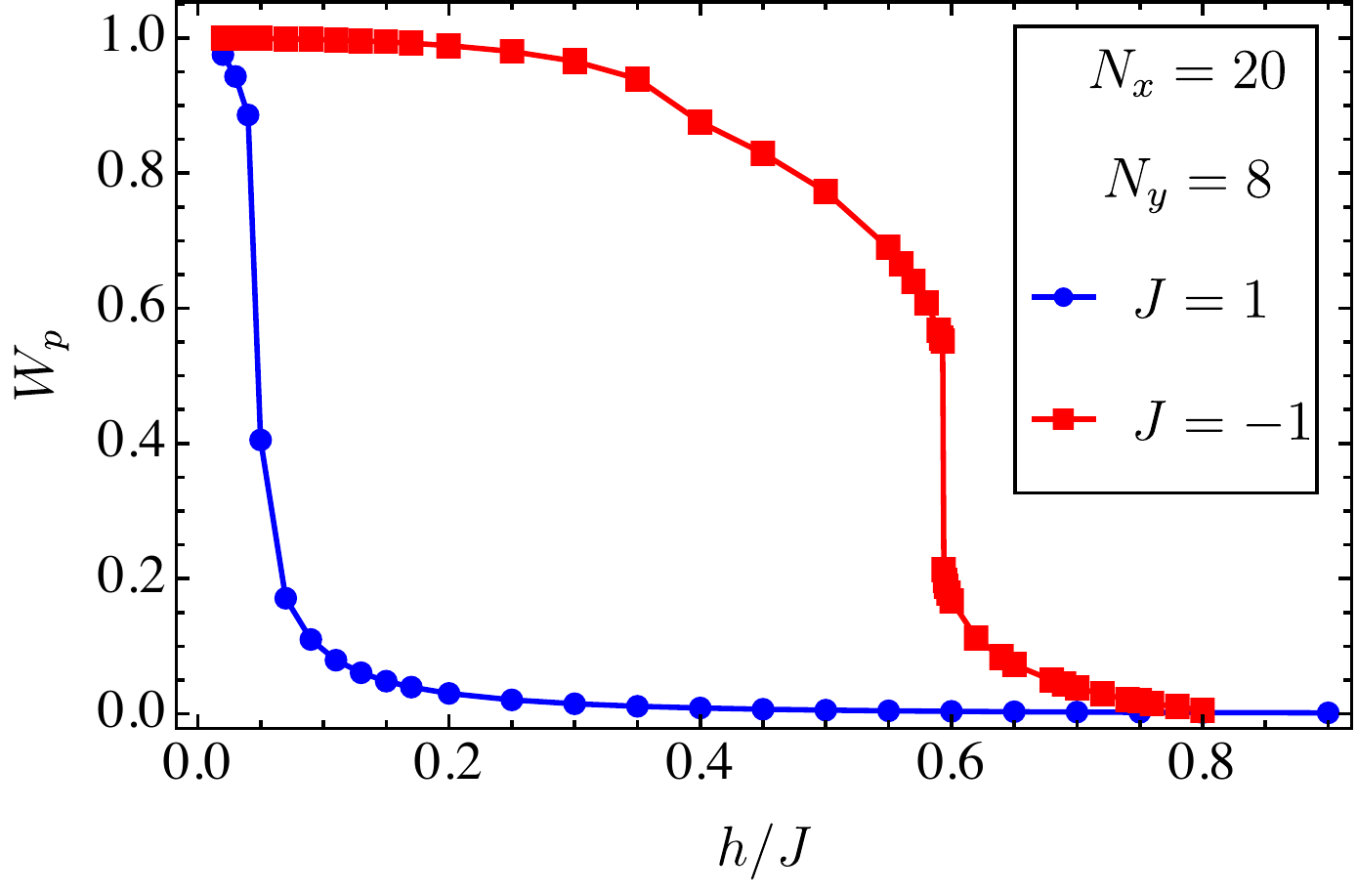}
\caption{\label{fig:flux_n} The expectation value of plaquette flux ($W_p$) for the Kitaev model with a Zeeman field in $z$ direction is shown for ferromagnetic ($J=1$, solid circles) and antiferromagnetic ($J=-1$, solid squares) cases using the finite MPS DMRG technique. The field at the transition for the AFM case is approximately $ 0.6|J|.$ }
\end{figure}

Figure \ref{fig:flux_n} shows a numerical calculation of $W_p$ for the ferromagnetic $(J>0)$ and antiferromagnetic ($J<0$) cases as a function of $(h/|J|)$ using the finite MPS DMRG technique. The maximum truncation error is $~10^{-10},$ and the calculation is for a system size of $N=20\times 8$. $W_p$ for the antiferromagnetic case sharply falls from $W_p = 1$ expected for the Kitaev phase at low fields at $h/|J| \approx 0.6$. This is to be compared with our prediction of $h/|J|\approx 0.52$ from Eq. \ref{eq:transitionpoint}. For the ferromagnetic case, the critical field is smaller, consistent with our prediction, but we do not have a quantitative estimate because our perturbation treatment does not work at low fields. 

\begin{figure}[t]
\centering
\includegraphics[width=\columnwidth]{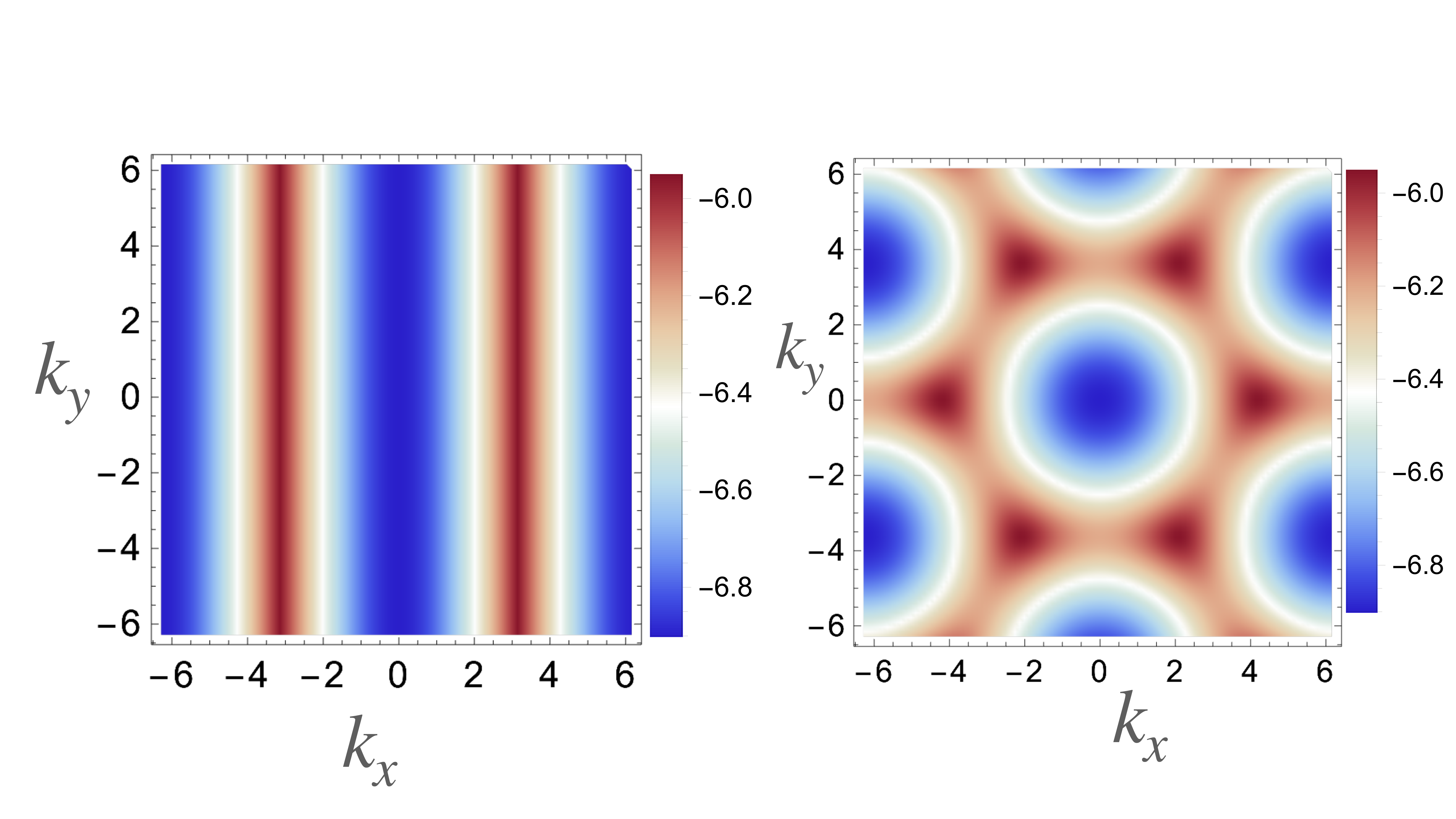}
\caption{\label{fig:1D} The contour plots; $(a)$ and $(b)$ illustrate the spin-wave dispersion in the  Kitaev model for Zeeman fields in $(001)$ and $(111)$ direction respectively (for $J=0.5$ and $h=5$) using the spin-wave treatment of Ref. \cite{joshi2018topological}. For $(001)$ oriented Zeeman fields, the dispersion is one-dimensional along the $x$-$y$ link backbone. The dispersion is two-dimensional for the $(111)$ orientation of the Zeeman field.}
\end{figure}

Another important feature of our analysis is the effective dimensional reduction of the excitations at high fields. In this regime, we can neglect the $E_J$ term, as a result of which the Maxwell term, now consisting only of the electric field contribution, lives only on the $x$-$y$ link backbone, and the $z$ links drop out. This implies that the excitations at high fields do not disperse along the $z$-direction. 
Observe that in the effective action for the phase degrees of freedom in Eq. (\ref{eq:seff}), the link fields are all decoupled when the Josephson term is discarded in the high field limit. The phase excitations would correspond to a flat band. To obtain the dispersion, we need to consider higher order time derivative contributions at $O(J^2),$ and the leading contribution at $O(J^{4}).$ The latter includes terms involving neighboring $x$ and $y$ links.  We refer to the Appendix for details of the higher order expansion. These terms contribute the following to the effective action in Eq. (\ref{eq:seff}):
\begin{align}
\frac{J^2}{  (4h)^{5}}\left\{\bigg(\frac{\partial^2 \varphi^{x}_1}{\partial \tau^{2}}\bigg)^{2}+\bigg( \frac{\partial^2 \varphi^{y}_1}{\partial \tau^2}\bigg)^2+\frac{3}{2}J^2\frac{\partial \varphi^{x}_1}{\partial \tau}\frac{\partial \varphi^{y}_1}{\partial \tau}\right\}
\end{align}
This gives two 1D dispersing modes (unit cell consists of two links),
\begin{align}
 E(k_x)  = \pm \sqrt{(4h)^{2} \pm (3/2) J^{2}\cos{k_x}}.
 \label{eq:1ddispersion}
\end{align}
at a high energy corresponding to the Zeeman gap. These modes are associated with the collective motion of fermion pairs. The spin-waves, which correspond to particle-hole excitations in our treatment, would be associated with the $\varphi_2^{x(y)}$ modes. Such processes in our theory would appear at $O(J^4)$ in the electric field terms. Other possible ways to generate the particle-hole hopping processes at lower order would be to locally switch off the Zeeman field at one or more sites, or even add a small component to the Zeeman field along the spin $x$ or $y$-directions. This allows particle-hole processes to act in the ground state sector. Although neither of these high energy collective modes is of interest to us for the effective low energy theory (constructed for energy scales much less than the Zeeman scale), they serve to illustrate the 1D nature of the physics at high fields. To compare with this understanding, we have performed a spin-wave calculation (see Fig. \ref{fig:1D}) using the analysis of Ref. \cite{joshi2018topological} for the field in the  $(001),$ i.e., $z$ direction. The 1D nature of the spin-wave excitations is clearly observed. Such dimensional reduction does not occur (see Fig.\ref{fig:1D}b), for example for fields $h$ in the $(111)$ direction where most of the existing studies have focused on \cite{joshi2018topological}. 
We conclude with a count of the degrees of freedom in our model. The original spin model has $2^{N}$ states, where $N$ is the number of spins. There are $N/2$ hexagons, each associated with a plaquette Wilson loop of value $\pm 1.$ Additionally, the matter fields $\Delta_i$ are $Z_2$ degrees of freedom on every $z$-link, which accounts for the remaining $2^{N/2}$ degrees of freedom.
\section{Discussion}\label{sec:5}
 In summary, we  have developed a mutual CS formalism for anyons on lattices lacking face-vertex correspondence. The spin degrees of freedom were expressed in terms of JW fermions coupled to the lattice $Z_2$ gauge fields and a certain local combination of the dual ($U(1)$ or $Z_2$) gauge fields, which ensured correct spin exchange statistics on arbitrary 2D lattices. In the presence of fermionic matter, the theory is not invariant under gauge transformations of the dual fields because of the absence of vortex matter on the dual lattice sites. 
 Our CS gauge theory of for 2D quantum spin systems qualitatively differs from earlier works (see e.g. Ref. \cite{fradkin2015prb}) where the JW fermions couple only to the lattice gauge fields, and the CS term involves only one kind of gauge fields. In Ref. \cite{fradkin2015prb}, although a mutual CS term had appeared as an intermediate step in transforming the lattice CS theory to a dual CS theory; the dual CS fields introduced there only serve the role of a decoupling field for the original CS action, and are not central to implementing correct spin statistics. Such approaches do not permit construction of consistent lattice matter-CS theories for spin lattices that lack face-vertex correspondence.

 As an illustration, the formulation was used to obtain an effective CS field theory of the honeycomb Kitaev model subjected to a strong Zeeman field in the $z$-direction. The effective theory is that of a superfluid coupled to fluctuating gauge fields whose dynamics is governed by a mutual Maxwell-Chern-Simons theory. The field-tuned topological transition of the Kitaev model appears as a normal to superfluid phase transition in our description. 
 
 We briefly discuss the excitations in the normal phases going up to the normal-superfluid transition. At high fields, $E_J/E_C \equiv 0$ for the FM Kitaev case (since $\Delta=0$ at high fields here) and small for AFM Kitaev ($\Delta \neq 0$ nonzero but small), which means only the electric part of the Maxwell term is important, and the gauge fields do not propagate. At low fields such that $E_J/E_C > 1,$ the cosine ``Josephson'' term can be expanded in increasing powers of the plaquette flux, and to quadratic order in the gauge fields, the result is a Maxwell-Chern-Simons like theory with a massive photon with group velocity $c \sim \sqrt{E_J E_C}\sim J^{2} |\Delta| /h,$ and the ``mass'' of the propagating photon mode is $\kappa c.$   In the vicinity of the transition to the topologically ordered phase (i.e. $g \sim 1$), the photon group velocity scales as $h \sqrt{|\Delta|},$ which agrees with estimates of the vison hopping scale $t_{\text{vison}}\sim h$ obtained from perturbative expansion in the Kitaev limit \cite{joy2021dynamics,chen2023nature}. Since for any value of the field, $|\Delta|$ is generally larger for the AFM Kitaev model, the vison dispersion persists to higher Zeeman fields in the AFM Kitaev case \cite{zhu2018robust}. In either case (FM or AFM Kitaev), vison propagation requires us to be in the superfluid phase (i.e. $g\gtrsim 1$). In order to validate the effectiveness of our approach, we have calculated that the expectation value of plaquette flux $W_p$ (see Fig. \ref{fig:flux_n}) using finite MPS DMRG technique for the AFM case (J = -1) sharply falls at $h/|J| \approx 0.6,$ which is consistent with our estimate of  $h/|J| \approx 0.52$ (see Eq. \ref{eq:transitionpoint}).  Although the confined phase  ($E_J/E_{C}\lesssim1$)  and deconfined phase ($E_J/E_{C}>1$) resemble the toric code in a Zeeman field \cite{kitaev2003fault,fradkin1979phase,trebst2007breakdown}, the $E_{J}/E_{C}$ contains an additional factor $|\Delta|^2$ which is zero for the FM case for sufficiently high fields $h > 2J.$ The presence of the Higgs field $\Delta$ makes our model different from a pure gauge theory such as the toric code. There is no Coulomb phase either, due to the presence of the CS term. We also identified an interesting dimensional reduction (see Eq. \ref{eq:1ddispersion}) at high fields where the Josephson term is negligible, and the model essentially reduces to disconnected 1D chains along the $xy$ backbone. We also  showed this quasi-1D dispersion also appears in a spin-wave analysis of the same regime (see Appendix). 
 
 Similar $Z_2$ Maxwell-CS gauge theories have also been proposed for quantum critical high temperature superconductors \cite{senthil2000z} where the superconductor-insulator transition is of the confinement-deconfinement type like ours and associated with $Z_2$ symmetry breaking. The original problem in that case even has a larger $U(1)$ gauge symmetry, which turns out to be not relevant for the superconductor-insulator transition.   
 
 Owing to the perturbative nature of our treatment, we were unable to study the properties of the deconfined phase. The deconfined phase is described by a level-$2$ mutual Chern-Simons theory, while at high fields we have a level-$1$ theory. Unfortunately, we were not able to identify a route to renormalization of the level with decreasing magnetic field up to the $O(J^4)$ perturbation, although we do not rule out appearance of such terms at higher order. An alternate route is a perturbative study from the low-field regime, which is a work in progress. The idea is to begin with a mutual Chern-Simons theory of level-$2$ that corresponds to a 4-fold degenerate ground state of the Kitaev model, and study how with increasing Zeeman and other perturbations, the level of the Chern-Simons theory would get renormalized to the high field value. It is important that in the low-field regime, the bare Green function has both sublattice diagonal and sublattice off-diagonal elements.  
 
 We finally discuss the cases where the Zeeman field has nonzero components in other directions (apart from $z$,) or more generally, the perturbations involve an odd number of fermions, long string-like excitations cannot be avoided. This would be taken up in a future study.

\begin{acknowledgments}
The authors acknowledge support of the Department of Atomic Energy, Government of India, under Project Identification No. RTI 4002, and Department of Theoretical Physics, TIFR, for computational resources. JD and VT thank Shiraz Minwalla and Subir Sachdev for discussions on aspects of Chern-Simons theory, and Kedar Damle for pointing out relevant literature and reading the manuscript.
\end{acknowledgments}
\vspace{0.5cm}
\textit{Statement of author contributions}: JD and VT conceived the problem and developed the mutual Chern-Simons formalism for spin-$1/2$ systems on arbitrary 2D lattices. AM contributed to the theoretical understanding and AK performed the DMRG calculations. JD and VT wrote the paper with inputs from AM and AK.

\section*{Appendix}

\subsection*{Perturbative expansion for the effective phase action}
\begin{widetext}
Here we briefly describe how the leading order link and loop terms in the effective action in Eq. (\ref{eq:seff}) have been obtained. The $O(J^2)$ contribution is given by expanding the fermionic determinant $\text{tr}\ln G^{-1} = \text{tr}\ln G_{0}^{-1} + \text{tr}\ln[1+ G_{0}(T_{0}+T)]$ to quadratic order in $(T_0 + T).$ For example,
 \begin{align}
  \text{tr}(G_{0}TG_{0}T)    
 =\int d{\tau}d{\tau'} \sum_{i_A,j_B}G^{i_A}_{0}(\tau-\tau')T^{i_Aj_B}(\tau')G^{j_B}_{0}(\tau'-\tau)T^{j_Bi_A}(\tau),
 \label{eq:trace}
 \end{align}
where the non-vanishing  tunneling matrices $T$ along the $x$ and $y$ links respectively have the form (Eq. \ref{eq:T}),
  \begin{equation}
 T^{i_Aj_B}\!=\!\frac{J_{x}}{2}\!\!
 \begin{bmatrix}
  -e^{i\varphi^{x}_{2}} & -e^{-i\varphi^{x}_{1}} \\
  e^{i\varphi^{x}_{1}}  & e^{-i\varphi^{x}_{2}} \\
 \end{bmatrix},
 T^{j_Bi_A}\!=\!\frac{J_{x}}{2}\!\!
 \begin{bmatrix}
 -e^{-i\varphi^{x}_{2}} & e^{-i\varphi^{x}_{1}}\\
 -e^{i\varphi^{x}_{1}} & e^{i\varphi^{x}_{2}}  \\
 \end{bmatrix},
 T^{k_Al_B}\!=\!\frac{J_{y}}{2}\!\!
 \begin{bmatrix}
 -e^{i\varphi^{y}_{2}} & e^{-i\varphi^{y}_{1}} \\
  -e^{i\varphi^{y}_{1}}  & e^{-i\varphi^{y}_{2}} \\
 \end{bmatrix},
 T^{l_Bk_A}\!=\!\frac{J_{y}}{2}\!\!
 \begin{bmatrix}
 -e^{-i\varphi^{y}_{2}} & -e^{-i\varphi^{y}_{1}}  \\
 e^{i\varphi^{y}_{1}} & e^{i\varphi^{y}_{2}} \\
 \end{bmatrix}.
 \label{eq:tunnelingmatrix}
 \end{equation} 
  So each term in the Eq. \ref{eq:trace} is a $2\times 2$ matrix. Here $i_A\rightarrow j_B$ is a $x$ bond and $k_A\rightarrow l_B$ is a $y$ bond.  The $\varphi's$ are defined in Eq. \ref{eq:phi-def}.
 The bare Green functions are all local in position coordinates and have the form ${G^{i_A}_{0}}^{11}(\tau-\tau')=\frac{1}{\beta}
 \sum_{\omega_{m}} \frac{e^{-i\omega_{m}(\tau-\tau')}}{i\omega_{m}-\xi_{i_A}},$ where ${G^{i_A}_{0}}^{11}$ is the first diagonal element of ${G^{i_A}_{0}}.$ Thus 
 \begin{align}
   \text{tr}(G_{0}TG_{0}T) & = 4\!\! \int_{{\tau},{\tau'}}\!\!\sum_{i,j,\omega_{1},\omega_{2}}\!\! \frac{1}{\beta^{2}} \!
\times\!\! \Bigg[\frac{e^{-i\omega_{1}(\tau-\tau')}}{i\omega_{1}-\xi_{i_A}}\frac{e^{-i\omega_{2}(\tau'-\tau)}}{i\omega_{2}-\xi_{j_B}}T_{11}^{{j_Bi_A}}(\tau)T_{11}^{{i_Aj_B}}(\tau')
  +\frac{e^{-i\omega_{1}(\tau-\tau')}}{i\omega_{1}+\xi_{i_A}}\frac{e^{-i\omega_{2}(\tau'-\tau)}}{i\omega_{2}+\xi_{j_B}}T_{22}^{{j_Bi_A}}(\tau)T_{22}^{{i_Aj_B}}(\tau') \nonumber \\
  & +\frac{e^{-i\omega_{1}(\tau-\tau')}}{i\omega_{1}-\xi_{i_A}}\frac{e^{-i\omega_{2}(\tau'-\tau)}}{i\omega_{2}+\xi_{j_B}}T_{21}^{{j_Bi_A}}(\tau)T_{12}^{{i_Aj_B}}(\tau') 
 +\frac{e^{-i\omega_{1}(\tau-\tau')}}{i\omega_{1}+\xi_{i_A}}\frac{e^{-i\omega_{2}(\tau'-\tau)}}{i\omega_{2}-\xi_{j_B}}T_{12}^{{j_Bi_A}}(\tau)T_{21}^{{i_Aj_B}}(\tau')\Bigg]
 \label{eq:secondordertrace}
 \end{align}
 where  $T_{11}^{i_Aj_B}$ and $T_{22}^{i_Aj_B}$ are the  diagonal  components of the $2\times 2$ matrix $T^{i_Aj_B}$  and $T_{12}^{i_Aj_B}$ and $T_{21}^{i_Aj_B}$ are off diagonal and $\xi_{i_A}=\xi_{j_B}  \approx 2h$. 
 It is convenient to work with the Green functions in Euclidean time:
 \begin{align}
     \frac{1}{\beta}\sum_{\omega}\frac{e^{-i\omega(\tau-\tau')}}{i\omega-\xi_{i_A}} 
     =-[\theta(\tau-\tau')n_{F}(-\xi_{i_A})-\theta(\tau'-\tau)n_{F}(\xi_{i_A})]e^{-\xi_{i_A}(\tau-\tau')}
     \label{matsubarasum}
 \end{align}
 \iffalse
 \begin{equation*}
      \frac{1}{\beta}\sum_{\omega}\frac{e^{-i\omega(\tau'-\tau)}}{i\omega+\xi_{i_A}}
 \end{equation*}
 \begin{equation}      
      =-[\theta(\tau'-\tau)n_{F}(\xi_{i_A})-\theta(\tau-\tau')n_{F}(-\xi_{i_A})]e^{\xi_{i_A}(\tau'-\tau)}
 \end{equation}
 \fi
 where $n_{F}$ is Fermi-Dirac distribution and  $\theta(\tau-\tau')$ is the Heaviside step function with,
 \begin{eqnarray}
 \theta(\tau-\tau')=
          \begin{cases}
          1 & \text{for}\hspace{0.25in} \tau > \tau'\\
          0 & \text{for}\hspace{0.25in} \tau < \tau'\\
          \frac{1}{2} & \text{for}\hspace{0.25in} \tau = \tau'.
          \end{cases}
\end{eqnarray}  
Using these relations, the first two terms of the above trace give,
 \begin{align}
\text{tr}(G_{0}TG_{0}T) 
& = \int_{{\tau},{\tau'}} \sum_{i,j} 
  e^{-(\xi_{i_A}-\xi_{j_B})(\tau-\tau')} \nonumber \\ \times
  &  \Bigg\{\big{[} \theta(\tau-\tau')n_{F}(-\xi_{i_A})-\theta(\tau'-\tau)n_{F}(\xi_{i_A})\big{]}
   \times \big{[}\theta(\tau'-\tau)n_{F}(-\xi_{j_B})-\theta(\tau-\tau')n_{F}(\xi_{j_B})\big{]}
     \times T_{11}^{{j_Bi_A}}(\tau)T_{11}^{{i_Aj_B}}(\tau')\nonumber \\
    & +\big{[}\theta(\tau-\tau')n_{F}(\xi_{i_A})-\theta(\tau'-\tau)n_{F}(-\xi_{i_A})\big{]} \times \big{[}\theta(\tau'-\tau)n_{F}(\xi_{j_B})-\theta(\tau-\tau')n_{F}(-\xi_{j_B})\big{]}
 \times T_{22}^{{j_Bi_A}}(\tau)T_{22}^{{i_Aj_B}}(\tau')  
    \Bigg\}.
 \end{align}
This contribution is proportional to $n_{F}(-\xi_{i_A})n_{F}(\xi_{i_A}),$ which vanishes at zero temperature.
The only non-trivial terms are coming from the last two terms of the Eq. \ref{eq:secondordertrace} and $\text{tr}(G_{0}TG_{0}T)$ becomes
 \begin{align}
   \text{tr}(G_{0}TG_{0}T) &= -4\int   d{\tau}d{\tau'} \sum_{i,j} \theta(\tau-\tau')n_{F}(-\xi_{i_A})n_{F}(-\xi_{j_B})\times  e^{-(\xi_{i_A}+\xi_{j_B})(\tau-\tau')}T_{12}^{i_Aj_B}(\tau')T_{21}^{j_Bi_A}(\tau)\nonumber \\
 &  -4\int   d{\tau}d{\tau'} \sum_{i,j} \theta(\tau'-\tau)n_{F}(-\xi_{i_A})n_{F}(-\xi_{j_B})  \times e^{-(\xi_{i_A}+\xi_{j_B})(\tau'-\tau)}T_{21}^{i_Aj_B}(\tau')T_{12}^{j_Bi_A}(\tau). 
 \label{eq:app42}
 \end{align}
  We now make a change of variables, $\tau = \tau_{c}+\frac{\tau_{r}}{2}$ and  $\tau' = \tau_{c}-\frac{\tau_{r}}{2}$.  So
 \begin{align}
 \text{tr}(G_{0}TG_{0}T) & = -4\int^{\beta}_{0} d{\tau_{c}}   \int^{\beta}_{0}   d{\tau'_{r}} \sum_{i,j} n_{F}(-\xi_{i_A})n_{F}(-\xi_{j_B}) 
 \times e^{-(\xi_{i_A}+\xi_{j_B})(\tau_{r})}T_{12}^{i_Aj_B}(\tau_{c}-\frac{\tau_{r}}{2})T_{21}^{j_Bi_A}(\tau_{c}+\frac{\tau_{r}}{2})\nonumber \\ 
& -4\int^{\beta}_{0} d{\tau_{c}}  \int^{0}_{-\beta}   d{\tau'_{r}} \sum_{i,j} n_{F}(-\xi_{i_A})n_{F}(-\xi_{j_B})
 \times e^{(\xi_{i_A}+\xi_{j_B})(\tau_{r})}T_{21}^{i_Aj_B}(\tau_{c}-\frac{\tau_{r}}{2})T_{12}^{j_Bi_A}(\tau_{c}+\frac{\tau_{r}}{2}).   
 \label{eq:off diagonal}
\end{align}
  If we change $\tau_{r} \rightarrow -\tau_{r}$ in the second term of  RHS of Eq. \ref{eq:off diagonal} then we will get exactly same as the first term of RHS. The product of two tunneling  matrices can be written by Taylor series expansion, 
  \begin{equation}
    T_{12}^{i_Aj_B}(\tau_{c}-\frac{\tau_{r}}{2})T_{21}^{j_Bi_A}(\tau_{c}+\frac{\tau_{r}}{2})= \frac{J^{2}_{x}}{4} e^{i\varphi^{x}_{1}(\tau_{c}+\frac{\tau_{r}}{2}) }e^{-i\varphi^{x}_{1}(\tau_{c}-\frac{\tau_{r}}{2}) }= \frac{J^{2}_{x}}{4}[1+i\tau_{r}\frac{\partial \varphi^{x}_{1}}{\partial \tau_{c}}-\frac{\tau^{2}_{r}}{2}(\frac{\partial \varphi^{x}_{1}}{\partial \tau_{c}})^{2}+...].
    \label{eq:Taylor}
  \end{equation}
 Integrating out $\tau_r$ the second order trace in the $x$ bond of Kitaev honeycomb model takes the form,
\begin{align}
    \text{tr}(G_{0}TG_{0}T)= -2J^{2}_{x}\int^{\beta}_{0} d{\tau} \sum^{(i_A \xrightarrow{e} j_B)}_{x-\text{links}} n_{F}(-\xi_{i_A})n_{F}(-\xi_{j_B})  
    \times [I_{1}+iI_{2}\frac{\partial \varphi^{x}_{1}}{\partial \tau}-\frac{1}{2}I_{3}(\frac{\partial \varphi^{x}_{1}}{\partial \tau})^{2}] \hspace{0.1in}\text{with} \hspace{0.1in} \tau_c \equiv \tau \hspace{0.1in}\text{and} 
\end{align}
\begin{align}
I_{1} &=\frac{1}{(\xi_{i_A}+\xi_{j_B})}\Big{[}1-e^{-\beta(\xi_{i_A}+\xi_{j_B})}\Big{]} \hspace{0.1in} \text{and} \hspace{0.1in}  I_{2} =\frac{1}{(\xi_{i_A}+\xi_{j_B})^{2}}[1\!\!-\!e^{-\beta(\xi_{i_A}+\xi_{j_B})}\!-\!\beta (\xi_{i_A}+\xi_{j_B})e^{-\beta(\xi_{i_A}+\xi_{j_B})}], \nonumber \\  &  I_{3} =\frac{2}{(\xi_{i_A}+\xi_{j_B})^{3}}- e^{-\beta(\xi_{i_A}+\xi_{j_B})} \Big{[}\frac{\beta^{2}}{(\xi_{i_A}+\xi_{j_B})}+\frac{2\beta}{(\xi_{i_A}+\xi_{j_B})^{2}}+\frac{2}{(\xi_{i_A}+\xi_{j_B})^{3}}\Big{]}. 
  \end{align}
  Similarly the corresponding  trace for $y$ bond  leads to,
  \begin{align}
      \text{tr}(G_{0}TG_{0}T) = -2J^{2}_{y}\int^{\beta}_{0} d{\tau} \sum^{(i_A \xrightarrow{e} j_B)}_{y-\text{links}} n_{F}(-\xi_{i_A})n_{F}(-\xi_{j_B})\times [I_{1}+iI_{2}\frac{\partial \varphi^{x}_{1}}{\partial \tau}-\frac{1}{2}I_{3}(\frac{\partial \varphi^{y}_{1}}{\partial \tau})^{2}]  
  \end{align}
where $\varphi^{y}_{1}$ and $\varphi^{y}_{2}$ are now on the $y$ bond with  the same structures with $\varphi^{x}_{1}$ and $\varphi^{x}_{2}$ respectively. 
Following a similar procedure as above we obtain for the $z$ bond,
\begin{align}
   \text{tr}(G_{0}T_{0}G_{0}T_{0}) & = -64J^2_{z} \int^{\beta}_{0}d{\tau} \sum_{z-\text{bonds}} n_{F}(-\xi_{i_A})n_{F}(-\xi_{j_B}) \times  \Bigg\{I_{1}|\Delta_{i}|^2+I_{2}[\Delta_{i}\frac{\partial \Delta^{*}_{i}}{\partial \tau}+i|\Delta_{i}|^2 \frac{\partial \varphi^{i}}{\partial \tau}] \nonumber
 \end{align}
 \begin{align}
  & -\frac{1}{2}I_{3}[\frac{\partial \Delta_{i}}{\partial \tau}\frac{\partial \Delta^{*}_{i}}{\partial \tau}-i\Delta_{i}\frac{\partial \Delta^{*}_{i}}{\partial \tau}\frac{\partial \varphi^{i}}{\partial \tau}+i\Delta^{*}_{i}\frac{\partial \Delta_{i}}{\partial \tau}\frac{\partial \varphi^{i}}{\partial \tau}+|\Delta_{i}|^2(\frac{\partial \varphi^{i}}{\partial \tau})^2]  \Bigg\}, 
 \end{align}
$\varphi^{i}_{1}=\chi_{i_A}+\chi_{i_B}.$ The Taylor expansion also yields quartic and higher order derivatives of the phase fields. However in this gradient expansion, they are smaller by factors of $(\omega/4 h)^{2}.$ Likewise, one can expand to higher order in tunneling - successive higher orders are smaller by a factors of $(J/4h)^{2}.$ Such terms are not retained in our leading expansion for the link contribution to the phase action but will become necessary (see below) for a discussion of collective phase modes.
 
 The loop terms appear only at sixth order in the tunneling. This will give us the magnetic field term of the Maxwell theory shown in Eq. \ref{eq:wilson-6th}. For any hexagonal loop in the Kitaev honeycomb model with high Zeeman field in $z$ direction , the sixth order trace leads to the following result,
\begin{align}
-\frac{1}{6}\text {tr}(G_{0}TG_{0}T_{0}G_{0}TG_{0}TG_{0}T_{0}G_{0}T)
=\frac{8}{3} \int^{\beta}_{0} d\tau \frac{J^{2}_{x}J^{2}_{y}J^{2}_{z}}{(2h)^{5}}\Bigg\{\Delta_{j}\Delta^{*}_{l}e^{i\int_{{\hexagon}} (\vec{A}.\vec{dl}+\vec{B}.\vec{dl})}+h.c.\Bigg\}.
\end{align}
Note that the hexagonal loop contains two $z$-tunnellings parametrized by $T_0$ and four $x$-$y$ tunnellings matrices parametrized by $T.$ Unlike the link terms, since the path in the loop is not retraced, the leading contribution here does not involve any time derivatives. \\
 Now at high field the above integrals $I_{1},I_{2}, I_{3}$ are simplified and the product $n_{F}(-\xi_{i_A})n_{F}(-\xi_{j_B})$ is taken to be unity. Using this the  ultimate action of our theory is given by 
 \begin{align}
    S & =-\int^{\beta}_{0} d{\tau} \Bigg{[}iL_{\text{CS}}-\sum_{i}\{4J_{z}\Delta_{i}\Delta^{*}_{i}-J_{z}-h\} \Bigg{]}  -J^{2}_{x}\int^{\beta}_{0} d{\tau} \sum^{(i_A \xrightarrow{e} j_B)}_{x-\text{links}} [I_{1}+iI_{2}\frac{\partial \varphi^{x}_{1}}{\partial \tau}-\frac{1}{2}I_{3}(\frac{\partial \varphi^{x}_{1}}{\partial \tau})^{2}] \nonumber \\
   &   -J^{2}_{y}\int^{\beta}_{0} d{\tau} \sum^{(i_A \xrightarrow{e} j_B)}_{y-\text{links}} [I_{1}+iI_{2}\frac{\partial \varphi^{y}_{1}}{\partial \tau}-\frac{1}{2}I_{3}(\frac{\partial \varphi^{y}_{1}}{\partial \tau})^{2}] -32J^2_{z} \int^{\beta}_{0}d{\tau} \sum_{z-\text{bonds}} \Bigg\{I_{1}|\Delta_{i}|^2+I_{2}\Big{[}\Delta_{i}\frac{\partial \Delta^{*}_{i}}{\partial \tau}+i|\Delta_{i}|^2 \frac{\partial \varphi_{i}}{\partial \tau}\Big{]} \nonumber \\
   & -\frac{1}{2}I_{3}\Big{[}\frac{\partial \Delta_{i}}{\partial \tau}\frac{\partial \Delta^{*}_{i}}{\partial \tau}-i\Delta_{i}\frac{\partial \Delta^{*}_{i}}{\partial \tau}\frac{\partial \varphi_{i}}{\partial \tau}+i\Delta^{*}_{i}\frac{\partial \Delta_{i}}{\partial \tau}\frac{\partial \varphi_{i}}{\partial \tau}+|\Delta_{i}|^2(\frac{\partial \varphi_{i}}{\partial \tau})^2\Big{]}  \Bigg\}-\frac{8}{3} \int^{\beta}_{0} d\tau \frac{J^{2}_{x}J^{2}_{y}J^{2}_{z}}{(2h)^{5}}\Bigg\{\Delta_{j}\Delta^{*}_{l}e^{i\int_{{\hexagon}} (\vec{A}.\vec{dl}+\vec{B}.\vec{dl})}+h.c.\Bigg\}.\nonumber \\
  & =-\int^{\beta}_{0} d{\tau} \Bigg{[}iL_{\text{CS}}-\sum_{i}\{4J_{z}\Delta_{i}\Delta^{*}_{i}-J_{z}-h\} \Bigg{]} +\frac{J^{2}_{x}}{64h^3}\int^{\beta}_{0} d{\tau} \sum^{(i_A \xrightarrow{e} j_B)}_{x-\text{links}} [(\frac{\partial \varphi^{x}_{1}}{\partial \tau}-2ih)^2-12h^2]\nonumber \\
  & +\frac{J^{2}_{y}}{64h^3}\int^{\beta}_{0} d{\tau} \sum^{(i_A \xrightarrow{e} j_B)}_{y-\text{links}} [(\frac{\partial \varphi^{y}_{1}}{\partial \tau}-2ih)^2-12h^2] -32J^2_{z} \int^{\beta}_{0}d{\tau} \sum_{z-\text{bonds}} \Bigg\{I_{1}|\Delta_{i}|^2-I_{2}[\Delta^{*}(\partial_{\tau}-i\dot{\varphi^i})\Delta]-\frac{1}{2}I_{3}[|(\partial_{\tau}-i\dot{\varphi^i})\Delta|^{2} ] \Bigg\}\nonumber \\
 &-\frac{8}{3} \int^{\beta}_{0} d\tau \frac{J^{2}_{x}J^{2}_{y}J^{2}_{z}}{(2h)^{5}}\Bigg\{\Delta_{j}\Delta^{*}_{l}e^{i\int_{{\hexagon}} (\vec{A}.\vec{dl}+\vec{B}.\vec{dl})}+h.c.\Bigg\}.
\end{align}
\subsection*{Collective modes}

At high fields one can drop the loop terms in comparison to the link terms. The phase model then represents a set of decoupled rotors along the $x$-$y$ chains suggesting a one-dimensional character. We now show that these phase modes, although confined at low energies, ultimately begin to disperse at sufficiently high energies comparable to the Zeeman gap. For this we first need to consider higher order time derivatives at $O(J^2)$ in Eq. \ref{eq:Taylor} for $x$ and $y$ links: 
\begin{align}
  T_{12}^{i_Aj_B}(\tau_{c}-\frac{\tau_{r}}{2})T_{21}^{j_Bi_A}(\tau_{c}+\frac{\tau_{r}}{2})= \frac{J^{2}_{x}}{4} e^{i\varphi^{x}_{1}(\tau_{c}+\frac{\tau_{r}}{2}) }e^{-i\varphi^{x}_{1}(\tau_{c}-\frac{\tau_{r}}{2}) }= \frac{J^{2}_{x}}{4}[1+i\tau_{r}\frac{\partial \varphi^{x}_{1}}{\partial \tau_{c}}-\frac{\tau^{2}_{r}}{2}(\frac{\partial \varphi^{x}_{1}}{\partial \tau_{c}})^{2}+\frac{{\tau_r}^4}{24}\bigg(\frac{\partial^2 \varphi^{x}_1}{\partial \tau^{2}}\bigg)^{2}+...].    
\end{align}
After the $\tau_r,$ integration we  obtain the following additional contribution to the effective phase action for the same pair of links: 
 \begin{align}
\frac{J^2}{(4h)^5} \Big[\bigg(\frac{\partial^2 \varphi^{x}_1}{\partial \tau^{2}}\bigg)^{2}+\bigg( \frac{\partial^2 \varphi^{y}_1}{\partial \tau^2}\bigg)^2\Big],
 \end{align}
where $\tau=\tau_c.$ At this point, there is still no dispersion. We next calculate the fourth order tunnelling contribution associated with a pair of $x$ and  $y$ links sharing a vertex:
\begin{align}
  \int d{\tau_1}d{\tau_2}d{\tau_3}d{\tau_4}  \sum_{j_A,i_B,k_B}G^{iB}_{0}(\tau_1-\tau_2)T^{i_Bj_A}(\tau_2)G^{j_A}_{0}(\tau_2-\tau_3)T^{j_Ak_B}(\tau_3)G^{k_B}_{0}(\tau_3-\tau_4)T^{k_Bj_A}(\tau_4)G^{j_A}_{0}(\tau_4-\tau_1)T^{j_Ai_B}(\tau_1),
 \label{eq:trace_4}
 \end{align}
 where $j_A\rightarrow i_B$  and $j_A\rightarrow k_B$ are  $x$ bond  and $y$ bond respectively. We make a linear transformation on the imaginary time coordinates in terms of the average time $\tau_c$ and three relative coordinates $\tau_{r_1},\tau_{r_2},\tau_{r_3}$ as shown below;
 \begin{equation}
 \begin{bmatrix}
 \tau_1\\
 \tau_2\\
 \tau_3\\
 \tau_4\\
 \end{bmatrix} , 
=
 \begin{bmatrix}
 1 & \frac{3}{4}& \frac{1}{2} & \frac{1}{4} \\ 
 1 & -\frac{1}{4}& \frac{1}{2} & \frac{1}{4} \\ 
  1 & -\frac{1}{4}& -\frac{1}{2} & \frac{1}{4} \\ 
  1 & -\frac{1}{4}& -\frac{1}{2} & -\frac{3}{4}  \\ 
 \end{bmatrix}
 \begin{bmatrix}
 \tau_c\\
 \tau_{r_1}\\
 \tau_{r_2}\\
 \tau_{r_3}\\
 \end{bmatrix}.
 \label{relative}
 \end{equation}
 Now using Eq. \ref{eq:tunnelingmatrix} and Eq. \ref{matsubarasum} above, and making a Taylor expansion of the tunneling terms around $\tau,$ and finally integrating out the relative time coordinates, we obtain the following contribution to the effective Lagrangian at $O(J^4)$ for a pair of links on the $xy$-chains sharing a vertex:
 \begin{align}
 C\frac{J^4}{ (4 h)^5}\frac{\partial \varphi^{x}_1}{\partial \tau}\frac{\partial \varphi^{y}_1}{\partial \tau},
 \end{align}
 where $C = 3/2,$ an $O(1)$ number. 
Collecting these additional contributions together with Eq. \ref{eq:seff}, the Lagrangian for the phase modes on a pair of neighboring links has the quadratic part,
 \begin{align}
   \mathcal{L}[\varphi_{1}] & = \frac{J^2}{6 4 h^{3}}\left[\left(\frac{\partial \varphi^{x}_1}{\partial\tau}\right)^{2} +\left(\frac{\partial \varphi^{y}_1}{\partial\tau}\right)^{2}+ \frac{1}{ {(4h)}^{2}}\left\{\bigg(\frac{\partial^2 \varphi^{x}_1}{\partial \tau^{2}}\bigg)^{2}+\bigg( \frac{\partial^2 \varphi^{y}_1}{\partial \tau^2}\bigg)^2+\frac{3}{2}J^2\frac{\partial \varphi^{x}_1}{\partial \tau}\frac{\partial \varphi^{y}_1}{\partial \tau}\right\}\right]
 \end{align}
 gives us the  four 1D dispersing modes whose energies are given by $E(k_x)  = \pm \sqrt{(4h)^{2} \pm (3/2)J^{2}\cos{k_x}}.$

 For the spin-wave dispersion that is associated with $\varphi^{x/y}_2$ modes, we have almost the same type of  behaviour except that the Zeeman gap is $2h.$ These $\varphi^{x/y}_2$ first appear at $O(J^4)$ and we have to go to an even higher sixth order to get the dispersion of these modes. Physically, the $O(J^4)$ contribution here comes from two particle-hole hopping terms associated with the same link - the total energy of this pair is thus $4h.$  The collective modes studied here are clearly one-dimensional at high fields.  
\end{widetext}

\newpage
\bibliographystyle{apsrev4-2}
%\bibliography{references}

\begin{thebibliography}{41}%
\makeatletter
\providecommand \@ifxundefined [1]{%
 \@ifx{#1\undefined}
}%
\providecommand \@ifnum [1]{%
 \ifnum #1\expandafter \@firstoftwo
 \else \expandafter \@secondoftwo
 \fi
}%
\providecommand \@ifx [1]{%
 \ifx #1\expandafter \@firstoftwo
 \else \expandafter \@secondoftwo
 \fi
}%
\providecommand \natexlab [1]{#1}%
\providecommand \enquote  [1]{``#1''}%
\providecommand \bibnamefont  [1]{#1}%
\providecommand \bibfnamefont [1]{#1}%
\providecommand \citenamefont [1]{#1}%
\providecommand \href@noop [0]{\@secondoftwo}%
\providecommand \href [0]{\begingroup \@sanitize@url \@href}%
\providecommand \@href[1]{\@@startlink{#1}\@@href}%
\providecommand \@@href[1]{\endgroup#1\@@endlink}%
\providecommand \@sanitize@url [0]{\catcode `\\12\catcode `\$12\catcode
  `\&12\catcode `\#12\catcode `\^12\catcode `\_12\catcode `\%12\relax}%
\providecommand \@@startlink[1]{}%
\providecommand \@@endlink[0]{}%
\providecommand \url  [0]{\begingroup\@sanitize@url \@url }%
\providecommand \@url [1]{\endgroup\@href {#1}{\urlprefix }}%
\providecommand \urlprefix  [0]{URL }%
\providecommand \Eprint [0]{\href }%
\providecommand \doibase [0]{https://doi.org/}%
\providecommand \selectlanguage [0]{\@gobble}%
\providecommand \bibinfo  [0]{\@secondoftwo}%
\providecommand \bibfield  [0]{\@secondoftwo}%
\providecommand \translation [1]{[#1]}%
\providecommand \BibitemOpen [0]{}%
\providecommand \bibitemStop [0]{}%
\providecommand \bibitemNoStop [0]{.\EOS\space}%
\providecommand \EOS [0]{\spacefactor3000\relax}%
\providecommand \BibitemShut  [1]{\csname bibitem#1\endcsname}%
\let\auto@bib@innerbib\@empty
%</preamble>
\bibitem [{\citenamefont {Holstein}\ and\ \citenamefont
  {Primakoff}(1940)}]{holstein1940field}%
  \BibitemOpen
  \bibfield  {author} {\bibinfo {author} {\bibfnamefont {T.}~\bibnamefont
  {Holstein}}\ and\ \bibinfo {author} {\bibfnamefont {H.}~\bibnamefont
  {Primakoff}},\ }\bibfield  {title} {\bibinfo {title} {Field dependence of the
  intrinsic domain magnetization of a ferromagnet},\ }\href
  {https://doi.org/10.1103/PhysRev.58.1098} {\bibfield  {journal} {\bibinfo
  {journal} {Phys. Rev.}\ }\textbf {\bibinfo {volume} {58}},\ \bibinfo {pages}
  {1098} (\bibinfo {year} {1940})}\BibitemShut {NoStop}%
\bibitem [{\citenamefont {Wen}(2002)}]{wen2002quantum}%
  \BibitemOpen
  \bibfield  {author} {\bibinfo {author} {\bibfnamefont {X.-G.}\ \bibnamefont
  {Wen}},\ }\bibfield  {title} {\bibinfo {title} {Quantum orders and symmetric
  spin liquids},\ }\href {https://doi.org/10.1103/PhysRevB.65.165113}
  {\bibfield  {journal} {\bibinfo  {journal} {Phys. Rev. B}\ }\textbf {\bibinfo
  {volume} {65}},\ \bibinfo {pages} {165113} (\bibinfo {year}
  {2002})}\BibitemShut {NoStop}%
\bibitem [{\citenamefont {Teng}\ \emph {et~al.}(2020)\citenamefont {Teng},
  \citenamefont {Zhang}, \citenamefont {Samajdar}, \citenamefont {Scheurer},\
  and\ \citenamefont {Sachdev}}]{teng2020unquantized}%
  \BibitemOpen
  \bibfield  {author} {\bibinfo {author} {\bibfnamefont {Y.}~\bibnamefont
  {Teng}}, \bibinfo {author} {\bibfnamefont {Y.}~\bibnamefont {Zhang}},
  \bibinfo {author} {\bibfnamefont {R.}~\bibnamefont {Samajdar}}, \bibinfo
  {author} {\bibfnamefont {M.~S.}\ \bibnamefont {Scheurer}},\ and\ \bibinfo
  {author} {\bibfnamefont {S.}~\bibnamefont {Sachdev}},\ }\bibfield  {title}
  {\bibinfo {title} {Unquantized thermal hall effect in quantum spin liquids
  with spinon fermi surfaces},\ }\href
  {https://doi.org/10.1103/PhysRevResearch.2.033283} {\bibfield  {journal}
  {\bibinfo  {journal} {Phys. Rev. Res.}\ }\textbf {\bibinfo {volume} {2}},\
  \bibinfo {pages} {033283} (\bibinfo {year} {2020})}\BibitemShut {NoStop}%
\bibitem [{\citenamefont {Wang}\ and\ \citenamefont
  {Vishwanath}(2006)}]{wang2006spin}%
  \BibitemOpen
  \bibfield  {author} {\bibinfo {author} {\bibfnamefont {F.}~\bibnamefont
  {Wang}}\ and\ \bibinfo {author} {\bibfnamefont {A.}~\bibnamefont
  {Vishwanath}},\ }\bibfield  {title} {\bibinfo {title} {Spin-liquid states on
  the triangular and kagom{\'e} lattices: A projective-symmetry-group analysis
  of schwinger boson states},\ }\href
  {https://doi.org/10.1103/PhysRevB.74.174423} {\bibfield  {journal} {\bibinfo
  {journal} {Phys. Rev. B}\ }\textbf {\bibinfo {volume} {74}},\ \bibinfo
  {pages} {174423} (\bibinfo {year} {2006})}\BibitemShut {NoStop}%
\bibitem [{\citenamefont {Knolle}\ \emph {et~al.}(2018)\citenamefont {Knolle},
  \citenamefont {Bhattacharjee},\ and\ \citenamefont
  {Moessner}}]{knolle2018dynamics}%
  \BibitemOpen
  \bibfield  {author} {\bibinfo {author} {\bibfnamefont {J.}~\bibnamefont
  {Knolle}}, \bibinfo {author} {\bibfnamefont {S.}~\bibnamefont
  {Bhattacharjee}},\ and\ \bibinfo {author} {\bibfnamefont {R.}~\bibnamefont
  {Moessner}},\ }\bibfield  {title} {\bibinfo {title} {Dynamics of a quantum
  spin liquid beyond integrability: The kitaev-heisenberg-$\gamma$ model in an
  augmented parton mean-field theory},\ }\href
  {https://doi.org/10.1103/PhysRevB.97.134432} {\bibfield  {journal} {\bibinfo
  {journal} {Phys. Rev. B}\ }\textbf {\bibinfo {volume} {97}},\ \bibinfo
  {pages} {134432} (\bibinfo {year} {2018})}\BibitemShut {NoStop}%
\bibitem [{\citenamefont {Kitaev}\ and\ \citenamefont
  {Laumann}(2010)}]{kitaev2010topological}%
  \BibitemOpen
  \bibfield  {author} {\bibinfo {author} {\bibfnamefont {A.}~\bibnamefont
  {Kitaev}}\ and\ \bibinfo {author} {\bibfnamefont {C.}~\bibnamefont
  {Laumann}},\ }\bibfield  {title} {\bibinfo {title} {Topological phases and
  quantum computation},\ }\href@noop {} {\bibfield  {journal} {\bibinfo
  {journal} {Exact methods in low-dimensional statistical physics and quantum
  computing, Lecture Notes of the Les Houches Summer School}\ ,\ \bibinfo
  {pages} {101}} (\bibinfo {year} {2010})}\BibitemShut {NoStop}%
\bibitem [{\citenamefont {Kitaev}(2006)}]{kitaev2006anyons}%
  \BibitemOpen
  \bibfield  {author} {\bibinfo {author} {\bibfnamefont {A.}~\bibnamefont
  {Kitaev}},\ }\bibfield  {title} {\bibinfo {title} {Anyons in an exactly
  solved model and beyond},\ }\href {https://doi.org/10.1016/j.aop.2005.10.005}
  {\bibfield  {journal} {\bibinfo  {journal} {Ann. Phys. (N. Y.)}\ }\textbf
  {\bibinfo {volume} {321}},\ \bibinfo {pages} {2} (\bibinfo {year}
  {2006})}\BibitemShut {NoStop}%
\bibitem [{\citenamefont {Fradkin}(1989)}]{fradkin1989prl}%
  \BibitemOpen
  \bibfield  {author} {\bibinfo {author} {\bibfnamefont {E.}~\bibnamefont
  {Fradkin}},\ }\bibfield  {title} {\bibinfo {title} {Jordan-wigner
  transformation for quantum-spin systems in two dimensions and fractional
  statistics},\ }\href {https://doi.org/10.1103/PhysRevLett.63.322} {\bibfield
  {journal} {\bibinfo  {journal} {Phys. Rev. Lett.}\ }\textbf {\bibinfo
  {volume} {63}},\ \bibinfo {pages} {322} (\bibinfo {year} {1989})}\BibitemShut
  {NoStop}%
\bibitem [{\citenamefont {Wang}(1991)}]{wang1991ground}%
  \BibitemOpen
  \bibfield  {author} {\bibinfo {author} {\bibfnamefont {Y.}~\bibnamefont
  {Wang}},\ }\bibfield  {title} {\bibinfo {title} {Ground state of the
  two-dimensional antiferromagnetic heisenberg model studied using an extended
  wigner-jordon transformation},\ }\href
  {https://doi.org/10.1103/PhysRevB.43.3786} {\bibfield  {journal} {\bibinfo
  {journal} {Phys. Rev. B}\ }\textbf {\bibinfo {volume} {43}},\ \bibinfo
  {pages} {3786} (\bibinfo {year} {1991})}\BibitemShut {NoStop}%
\bibitem [{\citenamefont {Azzouz}(1993)}]{azzouz1993interchain}%
  \BibitemOpen
  \bibfield  {author} {\bibinfo {author} {\bibfnamefont {M.}~\bibnamefont
  {Azzouz}},\ }\bibfield  {title} {\bibinfo {title} {Interchain-coupling effect
  on the one-dimensional spin-1/2 antiferromagnetic heisenberg model},\ }\href
  {https://doi.org/10.1103/PhysRevB.48.6136} {\bibfield  {journal} {\bibinfo
  {journal} {Phys. Rev. B}\ }\textbf {\bibinfo {volume} {48}},\ \bibinfo
  {pages} {6136} (\bibinfo {year} {1993})}\BibitemShut {NoStop}%
\bibitem [{\citenamefont {Tong}(2016)}]{tong2016lectures}%
  \BibitemOpen
  \bibfield  {author} {\bibinfo {author} {\bibfnamefont {D.}~\bibnamefont
  {Tong}},\ }\bibfield  {title} {\bibinfo {title} {Lectures on the quantum hall
  effect},\ }\href@noop {} {\bibfield  {journal} {\bibinfo  {journal}
  {\href{https://arxiv.org/abs/1606.06687}{arXiv preprint arXiv:1606.06687}}\ }
  (\bibinfo {year} {2016})}\BibitemShut {NoStop}%
\bibitem [{\citenamefont {Xu}\ and\ \citenamefont
  {Sachdev}(2009)}]{subir2009prb}%
  \BibitemOpen
  \bibfield  {author} {\bibinfo {author} {\bibfnamefont {C.}~\bibnamefont
  {Xu}}\ and\ \bibinfo {author} {\bibfnamefont {S.}~\bibnamefont {Sachdev}},\
  }\bibfield  {title} {\bibinfo {title} {Global phase diagrams of frustrated
  quantum antiferromagnets in two dimensions: Doubled chern-simons theory},\
  }\href {https://doi.org/10.1103/PhysRevB.79.064405} {\bibfield  {journal}
  {\bibinfo  {journal} {Phys. Rev. B}\ }\textbf {\bibinfo {volume} {79}},\
  \bibinfo {pages} {064405} (\bibinfo {year} {2009})}\BibitemShut {NoStop}%
\bibitem [{\citenamefont {Kou}\ \emph {et~al.}(2008)\citenamefont {Kou},
  \citenamefont {Levin},\ and\ \citenamefont {Wen}}]{wen2008prb}%
  \BibitemOpen
  \bibfield  {author} {\bibinfo {author} {\bibfnamefont {S.-P.}\ \bibnamefont
  {Kou}}, \bibinfo {author} {\bibfnamefont {M.}~\bibnamefont {Levin}},\ and\
  \bibinfo {author} {\bibfnamefont {X.-G.}\ \bibnamefont {Wen}},\ }\bibfield
  {title} {\bibinfo {title} {Mutual chern-simons theory for z 2 topological
  order},\ }\href {https://doi.org/10.1103/PhysRevB.78.155134} {\bibfield
  {journal} {\bibinfo  {journal} {Phys. Rev. B}\ }\textbf {\bibinfo {volume}
  {78}},\ \bibinfo {pages} {155134} (\bibinfo {year} {2008})}\BibitemShut
  {NoStop}%
\bibitem [{\citenamefont {Dong}\ \emph {et~al.}(2008)\citenamefont {Dong},
  \citenamefont {Fradkin}, \citenamefont {Leigh},\ and\ \citenamefont
  {Nowling}}]{fradkin2008dhep}%
  \BibitemOpen
  \bibfield  {author} {\bibinfo {author} {\bibfnamefont {S.}~\bibnamefont
  {Dong}}, \bibinfo {author} {\bibfnamefont {E.}~\bibnamefont {Fradkin}},
  \bibinfo {author} {\bibfnamefont {R.~G.}\ \bibnamefont {Leigh}},\ and\
  \bibinfo {author} {\bibfnamefont {S.}~\bibnamefont {Nowling}},\ }\bibfield
  {title} {\bibinfo {title} {Topological entanglement entropy in chern-simons
  theories and quantum hall fluids},\ }\href
  {https://doi.org/10.1088/1126-6708/2008/05/016} {\bibfield  {journal}
  {\bibinfo  {journal} {J. High Energy Phys.}\ }\textbf {\bibinfo {volume}
  {2008}}\bibinfo  {number} { (05)},\ \bibinfo {pages} {016}}\BibitemShut
  {NoStop}%
\bibitem [{\citenamefont {Karch}\ and\ \citenamefont
  {Tong}(2016)}]{tong2016prx}%
  \BibitemOpen
\bibfield  {number} {  }\bibfield  {author} {\bibinfo {author} {\bibfnamefont
  {A.}~\bibnamefont {Karch}}\ and\ \bibinfo {author} {\bibfnamefont
  {D.}~\bibnamefont {Tong}},\ }\bibfield  {title} {\bibinfo {title}
  {Particle-vortex duality from 3d bosonization},\ }\href
  {https://doi.org/10.1103/PhysRevX.6.031043} {\bibfield  {journal} {\bibinfo
  {journal} {Phys. Rev. X}\ }\textbf {\bibinfo {volume} {6}},\ \bibinfo {pages}
  {031043} (\bibinfo {year} {2016})}\BibitemShut {NoStop}%
\bibitem [{\citenamefont {Sun}\ \emph {et~al.}(2015)\citenamefont {Sun},
  \citenamefont {Kumar},\ and\ \citenamefont {Fradkin}}]{fradkin2015prb}%
  \BibitemOpen
  \bibfield  {author} {\bibinfo {author} {\bibfnamefont {K.}~\bibnamefont
  {Sun}}, \bibinfo {author} {\bibfnamefont {K.}~\bibnamefont {Kumar}},\ and\
  \bibinfo {author} {\bibfnamefont {E.}~\bibnamefont {Fradkin}},\ }\bibfield
  {title} {\bibinfo {title} {Discretized abelian chern-simons gauge theory on
  arbitrary graphs},\ }\href {https://doi.org/10.1103/PhysRevB.92.115148}
  {\bibfield  {journal} {\bibinfo  {journal} {Phys. Rev. B}\ }\textbf {\bibinfo
  {volume} {92}},\ \bibinfo {pages} {115148} (\bibinfo {year}
  {2015})}\BibitemShut {NoStop}%
\bibitem [{\citenamefont {Sen}\ \emph {et~al.}(2000)\citenamefont {Sen},
  \citenamefont {Sen}, \citenamefont {Sexton},\ and\ \citenamefont
  {Adams}}]{sen2000pre}%
  \BibitemOpen
  \bibfield  {author} {\bibinfo {author} {\bibfnamefont {S.}~\bibnamefont
  {Sen}}, \bibinfo {author} {\bibfnamefont {S.}~\bibnamefont {Sen}}, \bibinfo
  {author} {\bibfnamefont {J.~C.}\ \bibnamefont {Sexton}},\ and\ \bibinfo
  {author} {\bibfnamefont {D.~H.}\ \bibnamefont {Adams}},\ }\bibfield  {title}
  {\bibinfo {title} {Geometric discretization scheme applied to the abelian
  chern-simons theory},\ }\href {https://doi.org/10.1103/PhysRevE.61.3174}
  {\bibfield  {journal} {\bibinfo  {journal} {Phys. Rev. E}\ }\textbf {\bibinfo
  {volume} {61}},\ \bibinfo {pages} {3174} (\bibinfo {year}
  {2000})}\BibitemShut {NoStop}%
\bibitem [{\citenamefont {Geraedts}\ and\ \citenamefont
  {Motrunich}(2012)}]{geraedts2012monte}%
  \BibitemOpen
  \bibfield  {author} {\bibinfo {author} {\bibfnamefont {S.~D.}\ \bibnamefont
  {Geraedts}}\ and\ \bibinfo {author} {\bibfnamefont {O.~I.}\ \bibnamefont
  {Motrunich}},\ }\bibfield  {title} {\bibinfo {title} {Monte carlo study of a
  u (1)$\times$ u (1) system with $\pi$-statistical interaction},\ }\href
  {https://doi.org/10.1103/PhysRevB.85.045114} {\bibfield  {journal} {\bibinfo
  {journal} {Phys. Rev. B}\ }\textbf {\bibinfo {volume} {85}},\ \bibinfo
  {pages} {045114} (\bibinfo {year} {2012})}\BibitemShut {NoStop}%
\bibitem [{\citenamefont {Kantor}\ and\ \citenamefont
  {Susskind}(1991)}]{susskind1991}%
  \BibitemOpen
  \bibfield  {author} {\bibinfo {author} {\bibfnamefont {R.}~\bibnamefont
  {Kantor}}\ and\ \bibinfo {author} {\bibfnamefont {L.}~\bibnamefont
  {Susskind}},\ }\bibfield  {title} {\bibinfo {title} {A lattice model of
  fractional statistics},\ }\href
  {https://doi.org/10.1016/0550-3213(91)90029-W} {\bibfield  {journal}
  {\bibinfo  {journal} {Int. J. Mod. Phys. B}\ }\textbf {\bibinfo {volume}
  {5}},\ \bibinfo {pages} {2701} (\bibinfo {year} {1991})}\BibitemShut
  {NoStop}%
\bibitem [{\citenamefont {Zhang}(2022)}]{zhang2021}%
  \BibitemOpen
  \bibfield  {author} {\bibinfo {author} {\bibfnamefont {B.}~\bibnamefont
  {Zhang}},\ }\bibfield  {title} {\bibinfo {title} {Abelian chern-simons gauge
  theory on the lattice},\ }\href {https://doi.org/10.1103/PhysRevD.105.014507}
  {\bibfield  {journal} {\bibinfo  {journal} {Phys. Rev. D}\ }\textbf {\bibinfo
  {volume} {105}},\ \bibinfo {pages} {014507} (\bibinfo {year}
  {2022})}\BibitemShut {NoStop}%
\bibitem [{\citenamefont {Banks}\ and\ \citenamefont
  {Zhang}(2022)}]{banks2021}%
  \BibitemOpen
  \bibfield  {author} {\bibinfo {author} {\bibfnamefont {T.}~\bibnamefont
  {Banks}}\ and\ \bibinfo {author} {\bibfnamefont {B.}~\bibnamefont {Zhang}},\
  }\bibfield  {title} {\bibinfo {title} {Lattice bf theory, dumbbells, and
  composite fermions},\ }\href
  {https://doi.org/https://doi.org/10.1016/j.nuclphysb.2022.115877} {\bibfield
  {journal} {\bibinfo  {journal} {Nucl. Phys. B}\ }\textbf {\bibinfo {volume}
  {981}},\ \bibinfo {pages} {115877} (\bibinfo {year} {2022})}\BibitemShut
  {NoStop}%
\bibitem [{\citenamefont {Sedrakyan}\ \emph {et~al.}(2020)\citenamefont
  {Sedrakyan}, \citenamefont {Moessner},\ and\ \citenamefont
  {Kamenev}}]{Kamenevprb2020}%
  \BibitemOpen
  \bibfield  {author} {\bibinfo {author} {\bibfnamefont {T.}~\bibnamefont
  {Sedrakyan}}, \bibinfo {author} {\bibfnamefont {R.}~\bibnamefont
  {Moessner}},\ and\ \bibinfo {author} {\bibfnamefont {A.}~\bibnamefont
  {Kamenev}},\ }\bibfield  {title} {\bibinfo {title} {Helical spin liquid in a
  triangular xxz magnet from chern-simons theory},\ }\href
  {https://doi.org/10.1103/PhysRevB.102.024430} {\bibfield  {journal} {\bibinfo
   {journal} {Phys. Rev. B}\ }\textbf {\bibinfo {volume} {102}},\ \bibinfo
  {pages} {024430} (\bibinfo {year} {2020})}\BibitemShut {NoStop}%
\bibitem [{\citenamefont {Sedrakyan}\ \emph {et~al.}(2017)\citenamefont
  {Sedrakyan}, \citenamefont {Galitski},\ and\ \citenamefont
  {Kamenev}}]{kamenev2017prb}%
  \BibitemOpen
  \bibfield  {author} {\bibinfo {author} {\bibfnamefont {T.~A.}\ \bibnamefont
  {Sedrakyan}}, \bibinfo {author} {\bibfnamefont {V.~M.}\ \bibnamefont
  {Galitski}},\ and\ \bibinfo {author} {\bibfnamefont {A.}~\bibnamefont
  {Kamenev}},\ }\bibfield  {title} {\bibinfo {title} {Topological spin ordering
  via chern-simons superconductivity},\ }\href
  {https://doi.org/10.1103/PhysRevB.95.094511} {\bibfield  {journal} {\bibinfo
  {journal} {Phys. Rev. B}\ }\textbf {\bibinfo {volume} {95}},\ \bibinfo
  {pages} {094511} (\bibinfo {year} {2017})}\BibitemShut {NoStop}%
\bibitem [{\citenamefont {Kumar}\ and\ \citenamefont
  {Tripathi}(2022)}]{kumar2022absence}%
  \BibitemOpen
  \bibfield  {author} {\bibinfo {author} {\bibfnamefont {A.}~\bibnamefont
  {Kumar}}\ and\ \bibinfo {author} {\bibfnamefont {V.}~\bibnamefont
  {Tripathi}},\ }\bibfield  {title} {\bibinfo {title} {Absence of thermal hall
  half-quantization by field-suppression of magnetic order in a
  kitaev-heisenberg model},\ }\href@noop {} {\bibfield  {journal} {\bibinfo
  {journal} {\href{https://arxiv.org/abs/2208.13708}{arXiv preprint
  arXiv:2208.13708}}\ } (\bibinfo {year} {2022})}\BibitemShut {NoStop}%
\bibitem [{\citenamefont {Zhu}\ \emph {et~al.}(2018)\citenamefont {Zhu},
  \citenamefont {Kimchi}, \citenamefont {Sheng},\ and\ \citenamefont
  {Fu}}]{zhu2018robust}%
  \BibitemOpen
  \bibfield  {author} {\bibinfo {author} {\bibfnamefont {Z.}~\bibnamefont
  {Zhu}}, \bibinfo {author} {\bibfnamefont {I.}~\bibnamefont {Kimchi}},
  \bibinfo {author} {\bibfnamefont {D.}~\bibnamefont {Sheng}},\ and\ \bibinfo
  {author} {\bibfnamefont {L.}~\bibnamefont {Fu}},\ }\bibfield  {title}
  {\bibinfo {title} {Robust non-abelian spin liquid and a possible intermediate
  phase in the antiferromagnetic kitaev model with magnetic field},\ }\href
  {https://doi.org/10.1103/PhysRevB.97.241110} {\bibfield  {journal} {\bibinfo
  {journal} {Phys. Rev. B}\ }\textbf {\bibinfo {volume} {97}},\ \bibinfo
  {pages} {241110} (\bibinfo {year} {2018})}\BibitemShut {NoStop}%
\bibitem [{\citenamefont {Gordon}\ \emph {et~al.}(2019)\citenamefont {Gordon},
  \citenamefont {Catuneanu}, \citenamefont {S{\o}rensen},\ and\ \citenamefont
  {Kee}}]{gordon2019theory}%
  \BibitemOpen
  \bibfield  {author} {\bibinfo {author} {\bibfnamefont {J.~S.}\ \bibnamefont
  {Gordon}}, \bibinfo {author} {\bibfnamefont {A.}~\bibnamefont {Catuneanu}},
  \bibinfo {author} {\bibfnamefont {E.~S.}\ \bibnamefont {S{\o}rensen}},\ and\
  \bibinfo {author} {\bibfnamefont {H.-Y.}\ \bibnamefont {Kee}},\ }\bibfield
  {title} {\bibinfo {title} {Theory of the field-revealed kitaev spin liquid},\
  }\href {https://doi.org/10.1038/s41467-019-10405-8} {\bibfield  {journal}
  {\bibinfo  {journal} {Nat. Commun.}\ }\textbf {\bibinfo {volume} {10}},\
  \bibinfo {pages} {1} (\bibinfo {year} {2019})}\BibitemShut {NoStop}%
\bibitem [{\citenamefont {Song}\ and\ \citenamefont
  {Senthil}(2022)}]{song2022translation}%
  \BibitemOpen
  \bibfield  {author} {\bibinfo {author} {\bibfnamefont {X.-Y.}\ \bibnamefont
  {Song}}\ and\ \bibinfo {author} {\bibfnamefont {T.}~\bibnamefont {Senthil}},\
  }\bibfield  {title} {\bibinfo {title} {Translation-enriched $ z\_2 $ spin
  liquids and topological vison bands: Possible application to $\alpha -
  rucl_3$},\ }\href@noop {} {\bibfield  {journal} {\bibinfo  {journal}
  {\href{https://arxiv.org/abs/2206.14197}{arXiv:2206.14197}}\ } (\bibinfo
  {year} {2022})}\BibitemShut {NoStop}%
\bibitem [{\citenamefont {Joy}\ and\ \citenamefont
  {Rosch}(2021)}]{joy2021dynamics}%
  \BibitemOpen
  \bibfield  {author} {\bibinfo {author} {\bibfnamefont {A.~P.}\ \bibnamefont
  {Joy}}\ and\ \bibinfo {author} {\bibfnamefont {A.}~\bibnamefont {Rosch}},\
  }\bibfield  {title} {\bibinfo {title} {Dynamics of visons and thermal hall
  effect in perturbed kitaev models},\ }\href@noop {} {\bibfield  {journal}
  {\bibinfo  {journal}
  {\href{https://arxiv.org/abs/2109.00250}{arXiv:2109.00250}}\ } (\bibinfo
  {year} {2021})}\BibitemShut {NoStop}%
\bibitem [{\citenamefont {Eliezer}\ and\ \citenamefont
  {Semenoff}(1992)}]{semenoff1992prb}%
  \BibitemOpen
  \bibfield  {author} {\bibinfo {author} {\bibfnamefont {D.}~\bibnamefont
  {Eliezer}}\ and\ \bibinfo {author} {\bibfnamefont {G.}~\bibnamefont
  {Semenoff}},\ }\bibfield  {title} {\bibinfo {title} {Intersection forms and
  the geometry of lattice chern-simons theory},\ }\href
  {https://doi.org/https://doi.org/10.1016/0370-2693(92)90168-4} {\bibfield
  {journal} {\bibinfo  {journal} {Phys. Lett. B}\ }\textbf {\bibinfo {volume}
  {286}},\ \bibinfo {pages} {118} (\bibinfo {year} {1992})}\BibitemShut
  {NoStop}%
\bibitem [{\citenamefont {Freedman}\ \emph {et~al.}(2004)\citenamefont
  {Freedman}, \citenamefont {Nayak}, \citenamefont {Shtengel}, \citenamefont
  {Walker},\ and\ \citenamefont {Wang}}]{freedman2004ap}%
  \BibitemOpen
  \bibfield  {author} {\bibinfo {author} {\bibfnamefont {M.}~\bibnamefont
  {Freedman}}, \bibinfo {author} {\bibfnamefont {C.}~\bibnamefont {Nayak}},
  \bibinfo {author} {\bibfnamefont {K.}~\bibnamefont {Shtengel}}, \bibinfo
  {author} {\bibfnamefont {K.}~\bibnamefont {Walker}},\ and\ \bibinfo {author}
  {\bibfnamefont {Z.}~\bibnamefont {Wang}},\ }\bibfield  {title} {\bibinfo
  {title} {A class of p, t-invariant topological phases of interacting
  electrons},\ }\href
  {https://doi.org/https://doi.org/10.1016/j.aop.2004.01.006} {\bibfield
  {journal} {\bibinfo  {journal} {Ann. Phys. (N. Y.)}\ }\textbf {\bibinfo
  {volume} {310}},\ \bibinfo {pages} {428} (\bibinfo {year}
  {2004})}\BibitemShut {NoStop}%
\bibitem [{\citenamefont {Derzhko}(2001)}]{derzhko2001jordan}%
  \BibitemOpen
  \bibfield  {author} {\bibinfo {author} {\bibfnamefont {O.}~\bibnamefont
  {Derzhko}},\ }\bibfield  {title} {\bibinfo {title} {Jordan-wigner
  fermionization for spin-1/2 systems in two dimensions: A brief review},\
  }\href@noop {} {\bibfield  {journal} {\bibinfo  {journal}
  {\href{https://arxiv.org/abs/cond-mat/0101188}{arXiv preprint
  cond-mat/0101188}}\ } (\bibinfo {year} {2001})}\BibitemShut {NoStop}%
\bibitem [{\citenamefont {Lopez}\ \emph {et~al.}(1994)\citenamefont {Lopez},
  \citenamefont {Rojo},\ and\ \citenamefont {Fradkin}}]{lopez1994chern}%
  \BibitemOpen
  \bibfield  {author} {\bibinfo {author} {\bibfnamefont {A.}~\bibnamefont
  {Lopez}}, \bibinfo {author} {\bibfnamefont {A.}~\bibnamefont {Rojo}},\ and\
  \bibinfo {author} {\bibfnamefont {E.}~\bibnamefont {Fradkin}},\ }\bibfield
  {title} {\bibinfo {title} {Chern-simons theory of the anisotropic quantum
  heisenberg antiferromagnet on a square lattice},\ }\href
  {https://doi.org/10.1103/PhysRevB.49.15139} {\bibfield  {journal} {\bibinfo
  {journal} {Phys. Rev. B}\ }\textbf {\bibinfo {volume} {49}},\ \bibinfo
  {pages} {15139} (\bibinfo {year} {1994})}\BibitemShut {NoStop}%
\bibitem [{\citenamefont {Kumar}\ \emph {et~al.}(2014)\citenamefont {Kumar},
  \citenamefont {Sun},\ and\ \citenamefont {Fradkin}}]{kumar2014chern}%
  \BibitemOpen
  \bibfield  {author} {\bibinfo {author} {\bibfnamefont {K.}~\bibnamefont
  {Kumar}}, \bibinfo {author} {\bibfnamefont {K.}~\bibnamefont {Sun}},\ and\
  \bibinfo {author} {\bibfnamefont {E.}~\bibnamefont {Fradkin}},\ }\bibfield
  {title} {\bibinfo {title} {Chern-simons theory of magnetization plateaus of
  the spin-1 2 quantum xxz heisenberg model on the kagome lattice},\ }\href
  {https://doi.org/10.1103/PhysRevB.90.174409} {\bibfield  {journal} {\bibinfo
  {journal} {Phys. Rev. B}\ }\textbf {\bibinfo {volume} {90}},\ \bibinfo
  {pages} {174409} (\bibinfo {year} {2014})}\BibitemShut {NoStop}%
\bibitem [{\citenamefont {Dyson}(1956)}]{dyson1956general}%
  \BibitemOpen
  \bibfield  {author} {\bibinfo {author} {\bibfnamefont {F.~J.}\ \bibnamefont
  {Dyson}},\ }\bibfield  {title} {\bibinfo {title} {General theory of spin-wave
  interactions},\ }\href {https://doi.org/10.1103/PhysRev.102.1217} {\bibfield
  {journal} {\bibinfo  {journal} {Phys. rev.}\ }\textbf {\bibinfo {volume}
  {102}},\ \bibinfo {pages} {1217} (\bibinfo {year} {1956})}\BibitemShut
  {NoStop}%
\bibitem [{\citenamefont {Oguchi}(1960)}]{oguchi1960theory}%
  \BibitemOpen
  \bibfield  {author} {\bibinfo {author} {\bibfnamefont {T.}~\bibnamefont
  {Oguchi}},\ }\bibfield  {title} {\bibinfo {title} {Theory of spin-wave
  interactions in ferro-and antiferromagnetism},\ }\href
  {https://doi.org/10.1103/PhysRev.117.117} {\bibfield  {journal} {\bibinfo
  {journal} {Phys. Rev.}\ }\textbf {\bibinfo {volume} {117}},\ \bibinfo {pages}
  {117} (\bibinfo {year} {1960})}\BibitemShut {NoStop}%
\bibitem [{\citenamefont {Joshi}(2018)}]{joshi2018topological}%
  \BibitemOpen
  \bibfield  {author} {\bibinfo {author} {\bibfnamefont {D.~G.}\ \bibnamefont
  {Joshi}},\ }\bibfield  {title} {\bibinfo {title} {Topological excitations in
  the ferromagnetic kitaev-heisenberg model},\ }\href
  {https://doi.org/10.1103/PhysRevB.98.060405} {\bibfield  {journal} {\bibinfo
  {journal} {Phys. Rev. B}\ }\textbf {\bibinfo {volume} {98}},\ \bibinfo
  {pages} {060405} (\bibinfo {year} {2018})}\BibitemShut {NoStop}%
\bibitem [{\citenamefont {Chen}\ and\ \citenamefont
  {Villadiego}(2023)}]{chen2023nature}%
  \BibitemOpen
  \bibfield  {author} {\bibinfo {author} {\bibfnamefont {C.}~\bibnamefont
  {Chen}}\ and\ \bibinfo {author} {\bibfnamefont {I.~S.}\ \bibnamefont
  {Villadiego}},\ }\bibfield  {title} {\bibinfo {title} {Nature of visons in
  the perturbed ferromagnetic and antiferromagnetic kitaev honeycomb models},\
  }\href {https://doi.org/10.1103/PhysRevB.107.045114} {\bibfield  {journal}
  {\bibinfo  {journal} {Phys. Rev. B}\ }\textbf {\bibinfo {volume} {107}},\
  \bibinfo {pages} {045114} (\bibinfo {year} {2023})}\BibitemShut {NoStop}%
\bibitem [{\citenamefont {Kitaev}(2003)}]{kitaev2003fault}%
  \BibitemOpen
  \bibfield  {author} {\bibinfo {author} {\bibfnamefont {A.~Y.}\ \bibnamefont
  {Kitaev}},\ }\bibfield  {title} {\bibinfo {title} {Fault-tolerant quantum
  computation by anyons},\ }\href
  {https://doi.org/10.1016/S0003-4916(02)00018-0} {\bibfield  {journal}
  {\bibinfo  {journal} {Ann. Phys. (N. Y.)}\ }\textbf {\bibinfo {volume}
  {303}},\ \bibinfo {pages} {2} (\bibinfo {year} {2003})}\BibitemShut {NoStop}%
\bibitem [{\citenamefont {Fradkin}\ and\ \citenamefont
  {Shenker}(1979)}]{fradkin1979phase}%
  \BibitemOpen
  \bibfield  {author} {\bibinfo {author} {\bibfnamefont {E.}~\bibnamefont
  {Fradkin}}\ and\ \bibinfo {author} {\bibfnamefont {S.~H.}\ \bibnamefont
  {Shenker}},\ }\bibfield  {title} {\bibinfo {title} {Phase diagrams of lattice
  gauge theories with higgs fields},\ }\href
  {https://doi.org/10.1103/PhysRevD.19.3682} {\bibfield  {journal} {\bibinfo
  {journal} {Phys. Rev. D}\ }\textbf {\bibinfo {volume} {19}},\ \bibinfo
  {pages} {3682} (\bibinfo {year} {1979})}\BibitemShut {NoStop}%
\bibitem [{\citenamefont {Trebst}\ \emph {et~al.}(2007)\citenamefont {Trebst},
  \citenamefont {Werner}, \citenamefont {Troyer}, \citenamefont {Shtengel},\
  and\ \citenamefont {Nayak}}]{trebst2007breakdown}%
  \BibitemOpen
  \bibfield  {author} {\bibinfo {author} {\bibfnamefont {S.}~\bibnamefont
  {Trebst}}, \bibinfo {author} {\bibfnamefont {P.}~\bibnamefont {Werner}},
  \bibinfo {author} {\bibfnamefont {M.}~\bibnamefont {Troyer}}, \bibinfo
  {author} {\bibfnamefont {K.}~\bibnamefont {Shtengel}},\ and\ \bibinfo
  {author} {\bibfnamefont {C.}~\bibnamefont {Nayak}},\ }\bibfield  {title}
  {\bibinfo {title} {Breakdown of a topological phase: Quantum phase transition
  in a loop gas model with tension},\ }\href
  {https://doi.org/10.1103/PhysRevLett.98.070602} {\bibfield  {journal}
  {\bibinfo  {journal} {Phys. Rev. lett.}\ }\textbf {\bibinfo {volume} {98}},\
  \bibinfo {pages} {070602} (\bibinfo {year} {2007})}\BibitemShut {NoStop}%
\bibitem [{\citenamefont {Senthil}\ and\ \citenamefont
  {Fisher}(2000)}]{senthil2000z}%
  \BibitemOpen
  \bibfield  {author} {\bibinfo {author} {\bibfnamefont {T.}~\bibnamefont
  {Senthil}}\ and\ \bibinfo {author} {\bibfnamefont {M.~P.}\ \bibnamefont
  {Fisher}},\ }\bibfield  {title} {\bibinfo {title} {Z 2 gauge theory of
  electron fractionalization in strongly correlated systems},\ }\href
  {https://doi.org/10.1103/PhysRevB.62.7850} {\bibfield  {journal} {\bibinfo
  {journal} {Phys. Rev. B}\ }\textbf {\bibinfo {volume} {62}},\ \bibinfo
  {pages} {7850} (\bibinfo {year} {2000})}\BibitemShut {NoStop}%
\end{thebibliography}%
%
\end{document}